\documentclass[
reprint,amsmath,amssymb,aps,
superscriptaddress,longbibliography,
prx,
floatfix,
nobalancelastpage,
]{revtex4-2}

\usepackage{graphicx}
\usepackage{dcolumn}
\usepackage{bm}
\usepackage{epstopdf}
\usepackage{float}
\usepackage{braket}
\usepackage{dsfont}
\usepackage{amsmath}
\usepackage{siunitx}
\usepackage[italicdiff]{physics}
\usepackage[T1]{fontenc}
\usepackage[dvipsnames]{xcolor}
\usepackage{color,soul}
\usepackage{lipsum}
\usepackage[colorlinks=true, allcolors=blue]{hyperref}
\usepackage[export]{adjustbox}

\graphicspath{{figures/}}

\begin{document}

\title{Mitigating Residual Exchange Coupling in Resonant Singlet-Triplet Qubits}

\author{Jiheng Duan}

\author{Fernando Torres-Leal}

\author{John M. Nichol}
\email{john.nichol@rochester.edu}
\affiliation{Department of Physics and Astronomy, University of Rochester, Rochester, NY, 14627 USA}
\affiliation{University of Rochester Center for Coherence and Quantum Science, Rochester, NY, 14627, USA}

\begin{abstract}
We propose methods to mitigate single- and two-qubit control errors due to residual exchange coupling in systems of exchange-coupled resonant singlet-triplet qubits. 
Commensurate driving, where the pulse length is an integer multiple of the drive period, can mitigate errors from residual intra-qubit exchange, including effects from counter rotating terms and off-axis rotations, as well as leakage errors during two-qubit operations. 
Residual inter-qubit exchange creates crosstalk errors that reduce single-qubit control fidelities. 
We show that using a single-spin coupler between two resonant singlet-triplet qubits can reduce this crosstalk error by an order of magnitude. Assuming perfect coupler state preparation and realistic charge and hyperfine noise, we predict that coupler-assisted two-qubit gate errors can be below $3\times10^{-3}$ for gate times as short as $66~\text{ns}$, even in the presence of residual exchange levels exceeding several hundred kHz.
Our results suggest the potential of utilizing coupler-based architectures for large scale fault-tolerant spin qubit processors based on resonant singlet-triplet qubits.
\end{abstract}


\maketitle

\section{Introduction} \label{sec:introduction}

Spin qubits in semiconductor quantum dots~\cite{burkard2023semiconductor} are promising physical qubits for fault-tolerant quantum processors~\cite{takeda2022quantum, van2022phase, hetenyi2024tailoring}, due to their long lifetimes~\cite{steinacker2025industry, watson2017atomically, yang2013spin, stano2022review}, long-range spin shuttling~\cite{foster2025dephasing, de2025high, xue2024si, van2024coherent, noiri2022shuttling, jadot2021distant, mills2019shuttling}, and the potential for high-density integration~\cite{ha2025two, weinstein2023universal, vandersypen2017interfacing, neyens2024probing, zwerver2022qubits, rao2025modular, borsoi2024shared, john2024two}.
Recent experimental advances have demonstrated both single- and two-qubit gate fidelities exceeding 99\% for single-spin-$1/2$ qubits~\cite{steinacker2025industry, xue2022quantum, noiri2022fast, mills2022two}, reaching the threshold required for error correction using the surface code~\cite{fowler2012surface, bonilla2021xzzx}. 
Encoding quantum information in the joint state of more than one spin enables low-frequency control~\cite{petta2005coherent, laird2010coherent, kim2014quantum, eng2015isotopically, blumoff2022fast, weinstein2023universal}, high-fidelity single-shot readout~\cite{blumoff2022fast, barthel2009rapid, nakajima2017robust, harvey2018high, takeda2024rapid, park2025single}, and encoding in decoherence-free subspaces~\cite{sun2024full, bluhm2011dephasing, friesen2017decoherence, chirolli2008decoherence}. 
Singlet-triplet qubits based on two spins exemplify these potential advantages.

In the presence of a large magnetic-field gradient, the natural basis states of the singlet-triplet qubit are $\left|\uparrow\downarrow\right\rangle$ and $\left|\downarrow\uparrow\right\rangle$, forming the ``flip-flop'' or ``resonant'' singlet-triplet (RST) qubit~\cite{klauser2006nuclear, tosi2017silicon}. High-fidelity single-qubit control via oscillating exchange couplings, as well as initialization and readout via Pauli spin-blockade methods, have been demonstrated in RST qubits~\cite{shulman2014suppressing,nichol2017high,orona2018readout,sigillito2019coherent,takeda2020resonantly,tsoukalas2026dressed}. 
The $\left|\uparrow\downarrow\right\rangle$, $\left|\downarrow\uparrow\right\rangle$ basis is also the natural basis in which to consider a two-qubit gate based on the ZZ-interaction between exchange-coupled singlet-triplet qubits~\cite{wardrop2014exchange}.
While capacitive coupling has been employed to generate longitudinal inter-qubit interactions~\cite{nichol2017high, shulman2012demonstration}, the exchange-based ZZ-interaction has not yet been conclusively demonstrated between singlet-triplet qubits~\cite{qiao2021floquet}. 

However, nearly all spin qubits that utilize exchange coupling must contend with ``residual exchange'', which results from the fact that exchange coupling between single electrons in separate dots is always positive~\cite{loss1998quantum}. In the context of RST qubits, this means that the basis states are not purely $\left|\uparrow\downarrow\right\rangle$ and $\left|\downarrow\uparrow\right\rangle$, and thus not orthogonal to the oscillating control field. In addition, this means that any oscillating exchange coupling is also accompanied by a non-zero time-averaged exchange. 
For two exchange-coupled RST qubits, the residual inter-qubit exchange creates a non-zero ZZ-interaction, causing quantum crosstalk errors in single-qubit gates on both qubits~\cite{xue2022quantum, heinz2024analysis}. 
This form of crosstalk, arising from the ZZ-interaction, is fundamentally different from microwave crosstalk between the driving electrodes of neighboring quantum dots~\cite{heinz2022crosstalk, heinz2021crosstalk, undseth2023nonlinear}.

Common approaches to mitigate effects of residual exchange include modifying the gate architecture or tuning to minimize the residual exchange. Typical residual exchange values using these approaches are on the order of a few hundred kilohertz, significantly suppressing quantum crosstalk~\cite{xue2022quantum, steinacker2025industry}. 
In these cases however, high-amplitude barrier-gate voltage pulses are required to turn on the exchange coupling between qubits.
Such high-voltage pulses can disrupt the linear-regime gate virtualization, as described by the crosstalk matrix that captures capacitive coupling between all metallic gates~\cite{volk2019loading, borsoi2024shared, mills2019computer, mills2019shuttling, rao2025modular, kelly2023capacitive}.
Previous studies have also suggested that such quantum crosstalk errors resulting from residual exchange can be suppressed by matching drive timings with Rabi periods and conditional precession frequencies~\cite{heinz2024analysis}. 
By carefully synchronizing these time scales, residual-exchange-induced coherent errors can interfere destructively and cancel out. 
This approach enables improved gate fidelity without requiring complete suppression of the residual exchange coupling at the cost of additional gate complexity. 

In this work, we propose a method to mitigate single-qubit errors in RST qubits from both residual intra- and inter-qubit exchange. 
By commensurately pulsing the intra-qubit exchange coupling at integer multiples of the single-qubit driving period and compensating phase shifts with virtual-Z rotations~\cite{mckay2017efficient}, fast single-qubit gates with low coherent error can be achieved even in the presence of large residual intra-qubit exchange. 
To mitigate quantum crosstalk errors, we propose a single-spin coupler between two RST qubits~\cite{mehl2014two} to mediate a superexchange interaction~\cite{qiao2021long} with large on-off ratios.

This paper is organized as follows. In Sec.~\ref{sec:direct_model} we discuss the Heisenberg exchange model of two coupled RST qubits in a one-dimensional spin chain~\cite{wardrop2014exchange}. 
Focusing on a single RST qubit, we derive the intra-qubit residual exchange-induced error under resonant driving without invoking the rotating-wave approximation (RWA) in Sec.~\ref{sec:intra res exchange}.
The commensurate drive conditions are derived from the leading-order (first-order) Magnus expansion of the time evolution operator for a $\pi/2$ rotation.
The inter-qubit residual exchange-induced quantum crosstalk is discussed in Sec.~\ref{sec:inter res exchange}.
Quasi-static hyperfine noise and $1/f$ charge noise~\cite{connors2019low, connors2022charge, ye2024characterization, freeman2016comparison} are included to benchmark the performance of both $X_{\pi/2}$ and CZ gates.
In Sec.~\ref{sec:inter res exchange}, we propose a single-spin coupler architecture and evaluate all gate operations under a consistent noise environment.

\section{System and Model} \label{sec:direct_model}

\begin{figure}[t]
{\includegraphics[width= 0.5\textwidth]{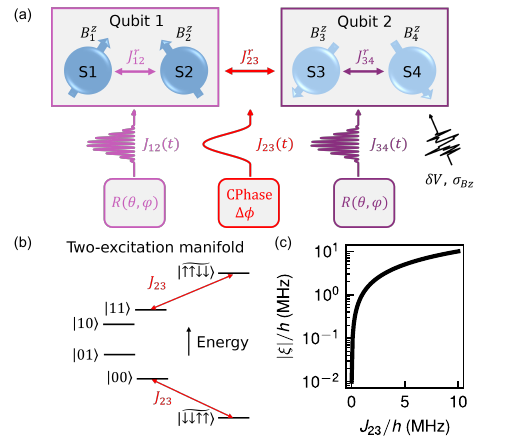}}
\caption{
(a) One-dimensional four-spin chain with spins $S1$ to $S4$. Two RST qubits are coupled though inter-qubit residual exchange $J_{23}^r$ with individual intra-qubit residual exchange couplings $J_{12}^r$ and $J_{34}^r$, respectively. Arbitrary single-qubit rotations can be realized by resonant intra-qubit exchange pulses together with virtual-Z gates. A CPhase gate can be achieved by pulsing inter-qubit exchange coupling. 
(b) Energy levels. The middle four levels and the remaining two levels span the computational and leakage subspaces, respectively. The inter-qubit exchange coupling $J_{23}$ couples states between the two subspaces, causing leakage errors. 
(c) ZZ-interaction $\zeta$ vs. inter-qubit exchange coupling $J_{23}$.
}
\label{fig:direct_sq_model}
\end{figure}

The RST qubit is constructed from the joint spin states of two electrons in a double quantum-dot in a large magnetic field gradient. 
In the absence of exchange coupling, the eigenstates of the system are  $\left\{ \left|\downarrow\downarrow\right\rangle,  \left|\downarrow\uparrow\right\rangle, \left|\uparrow\downarrow\right\rangle, \left|\uparrow\uparrow\right\rangle \right\}$.  In the presence of exchange coupling, the eigenstates of the system hybridize slightly, and we will refer to them in the following as $\left\{ |\widetilde{\downarrow\downarrow}\rangle,  |\widetilde{\downarrow\uparrow}\rangle, |\widetilde{\uparrow\downarrow}\rangle, |\widetilde{\uparrow\uparrow}\rangle\right\}$. 
The computational subspace of the RST qubit is spanned by the logical states $\left\{|0\rangle, |1\rangle \right\} = \left\{|\widetilde{\downarrow\uparrow}\rangle, |\widetilde{\uparrow\downarrow}\rangle \right\}$.  
Consider only the left RST qubit in Figure~\ref{fig:direct_sq_model}(a), decoupled from the second qubit.
The corresponding qubit Hamiltonian inside the logical subspace is
\begin{equation}
    H_0 = - \frac{1}{2} \hbar \sqrt{\Delta \omega_{12}^2 + \left(\frac{J_{12}^r}{\hbar}\right)^2} \sigma_z,
\end{equation}
where $\hbar$ is the reduced Planck constant, $\Delta\omega_{12}=\omega_{s1}-\omega_{s2}$ is the spin Larmor frequency difference between spins $S_1$ and $S_2$, $J_{12}^r$ is the residual inter-qubit exchange, and $\sigma_k$, $k\in\{x,y,z\}$ is the Pauli operator.

\begin{figure*}[t]
    {\includegraphics[width=\textwidth]{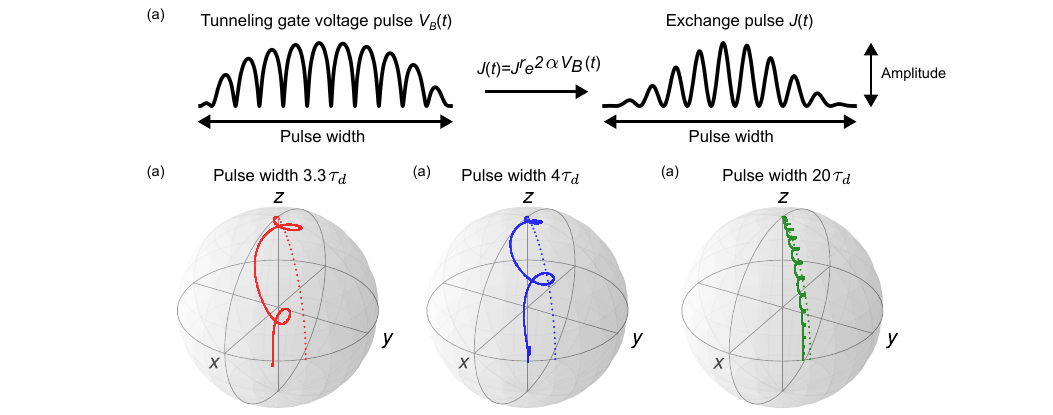}}
    \caption{
    (a) An inter-dot tunneling-gate voltage pulse with a logarithmic-cosine envelope can generate an exchange pulse with a cosine envelope. 
    (b)-(d) Trajectory on the Bloch sphere during a $\pi/2$ rotation with phase $\pi/4$, illustrated in solid lines. The corresponding dashed lines are the ideal rotations.
    The pulse width at each case is (b) $3.3\tau_d$, (c) $4\tau_d$, and (d) $20\tau_d$. An unfinished fast rotation for $3.3\tau_d$ is observed. 
    }
    \label{fig:direct_sq_pulse}
\end{figure*}

Single-qubit control is achieved by resonantly pulsing the intra-qubit exchange coupling, illustrated both in Fig.~\ref{fig:direct_sq_model}(a) and Fig.~\ref{fig:direct_sq_pulse}(a).
Modulating the carrier phase $\varphi$ enables the implementation of virtual-Z gates and arbitrary single-qubit rotations $R(\theta, \varphi)=\exp\left[ -i\theta(\cos\varphi\sigma_x - \sin\varphi\sigma_y)/2 \right]$, as demonstrated in  GaAs~\cite{nichol2017high}, Si~\cite{takeda2020resonantly}, and Ge~\cite{tsoukalas2026dressed}.
Because the exchange coupling between two electrons is non-negative, any control sequence that modulates the exchange also creates a non-zero residual exchange $J^r$. 
A non-zero residual exchange means that the quantization and control axes of the qubit are non-orthogonal and leads to pulse-dependent coherent errors, as discussed further in Sec.~\ref{sec:intra res exchange}.

Two-qubit gates are achieved by manipulating the exchange coupling between two RST qubits. 
Considering two exchange-coupled RST qubits in the presence of a large Zeeman gradient, illustrated in Fig.~\ref{fig:direct_sq_model}(a), the inter-qubit exchange coupling naturally creates a ZZ-interaction between two far-detuned spins. 
The energy levels of the two-excitation manifold are shown in Fig.~\ref{fig:direct_sq_model}(b) where the two-qubit computational subspace is spanned by
$\left\{|00\rangle, |01\rangle, |10\rangle, |11\rangle\right\}=
\left\{|\widetilde{\downarrow\uparrow\downarrow\uparrow}\rangle, 
|\widetilde{\downarrow\uparrow\uparrow\downarrow}\rangle, 
|\widetilde{\uparrow\downarrow\downarrow\uparrow}\rangle, 
|\widetilde{\uparrow\downarrow\uparrow\downarrow}\rangle\right\}$. 
Such a ZZ-interaction between spins from two RST qubits induces an effective ZZ-interaction in the two-qubit computational subspace. The strength of the coupling is defined as~\cite{zhao2020high}
\begin{equation} \label{eq:direct_zz}
    \zeta = E_{11} + E_{00} - E_{10} - E_{01},
\end{equation}
where $E_{kl}$ is the eigenenergy of the two-qubit eigenstate $|k,l\rangle$ for $k,l\in\{0,1\}$.
Figure~\ref{fig:direct_sq_model}(c) shows the ZZ-interaction against the inter-qubit exchange coupling $J_{23}$. 
The inter-qubit exchange coupling, however, also induces leakage out of the computational subspace. 
The leakage subspace is spanned by $\left\{
|\widetilde{\uparrow\uparrow\downarrow\downarrow}\rangle, 
|\widetilde{\downarrow\downarrow\uparrow\uparrow}\rangle
\right\}$ and is connected to the computational subspace through the inter-qubit exchange coupling $J_{23}$.
Methods such as adiabatic pulse shaping have been proposed to suppress leakage errors~\cite{wardrop2014exchange}.

In the following, we will explore the effects of residual exchange on RST qubits with analytical and numerical approaches. To numerically simulate the behavior of coupled RST qubits, we consider a complete four-spin chain with a Hamiltonian
\begin{equation}\label{eq:direct_Heisenberg}
H_\text{s} = -\frac{1}{2} \sum_{k=1}^4 \hbar\omega_{s,k}s_k^z + \frac{1}{4}\sum_{k=1}^3 J_{k,k+1}^r \vec{s}_k\cdot\vec{s}_{k+1}.
\end{equation}
Here $\hbar\omega_{s,k} = g_k{\mu_B}B_k^z$ is the Zeeman energy, $g_k$ is the gyromagnetic ratio, $\mu_B$ is the Bohr magneton, and $B_k^z$ $\left(k=1,...,4\right)$ is the local magnetic field in the z-direction, $s_k^\mu$ is the Pauli operator of the decoupled spins, and $\vec{s}_k = (s_k^x,s_k^y,s_k^z)$.
The residual exchange couplings between spins within the same qubit and those connecting different qubits are referred to as intra-qubit ($J_{12}^r$, $J_{34}^r$) and inter-qubit ($J_{23}^r$) residual exchange, respectively.

In our simulations we assume that exchange is controlled by pulsing the effective inter-dot tunnel barrier gate voltage $V_B(t)$.
It can be modeled as~\cite{pan2020resonant,xue2022quantum,walelign2024dynamically,steinacker2025industry} 
\begin{equation}\label{eq:voltage_2_exchange}
J(t) = J^r e^{2\alpha V_B(t)},
\end{equation}
where $J^r$ is the residual exchange coupling and $\alpha = 0.011~\text{mV}^{-1}$ is an empirical lever arm. We numerically simulate the evolution of the four-spin system under the influence of time-dependent exchange couplings using QuTiP~\cite{lambert2024qutip, jhd2024qusim}. 
Table~\ref{tab:spin_coupler_param} lists the simulation parameters we use. 

In our simulations, unless stated otherwise, we  also include the effects of both charge noise and hyperfine noise. 
We include charge noise by generating a noise time series with a $1/f$ power spectral density (PSD) 
\begin{equation}\label{eq:charge_noise_level}
S_{V_B}(f) = \frac{A_\mu^2\gamma^{-2}}{f},
\end{equation}
where $A_\mu = 1~\mu\text{eV}$ is the PSD amplitude at 1 Hz, and $\gamma=0.1~\text{eV}/\text{V}$ is the lever arm~\cite{connors2019low,yang2019achieving,ye2024characterization}. 
Details of the charge noise simulations are discussed in Appendix~\ref{apd:charge_noise}. We note that the voltage noise associated with the barrier becomes dominant at high exchange coupling as its sensitivity exponentially grows with the barrier voltage:
\begin{equation}\label{eq:chrg_noise_sensitivity}
    \frac{\partial J}{\partial V_B} = 2\alpha J^r e^{2\alpha V_B}.
\end{equation}
This fact has important implications for RST gate fidelities, as discussed further below. 

For spin $k$, the contribution from hyperfine noise is formulated as a stochastic offset on its Larmor frequency $\omega_{s,k} + \delta \omega_{s,k}$.
The hyperfine noise term $\delta\omega_{s,k}$ is considered as quasi-static and its amplitude is described by the standard deviation $\sigma_{Bz}$ of its Gaussian distribution $\delta\omega_{s,k}\sim\mathcal{N}(0,\sigma_{Bz}^2)$ with zero mean. 
Here, we use hyperfine noise with amplitude $\sigma_{Bz} = 2\pi\times0.05~\text{MHz}$, which corresponds to a simulated single-spin Ramsey coherence time $T_{2}^{*} \approx 2.37~\mu\text{s}$.
Further details about hyperfine noise are discussed in Appendix~\ref{apd:hyperfine_noise}. 
High-frequency noise channels that primarily induce depolarization (erasure) are captured by our simulated charge noise. However, we expect that their contribution is negligible for the gate timescales considered here~\cite{borjans2019single}. 

Practically, the RST qubits in our platform are usually exposed to large external spin-orbit coupling through the magnetic gradient.
Electrons in neighboring dots could couple to the same electrical fluctuators, resulting in both correlated exchange and Zeeman-energy noise~\cite{yoneda2023noise, rojas2026scaling, zou2024spatially, rojas2023spatial, rojas2025inferring, kobayashi2025charge}. In our simulations, we do not consider correlated noise and assume all noise channels are uncorrelated for benchmarking the control behavior. 
However, to explore effects from correlated charge noise, we perform additional simulations that consider a worst case scenario when all spins in the system are exposed to a common noise source (i.e., all spins suffer from correlated exchange and Zeeman energies noise) in Appendix~\ref{apd:correlated_noise}. 

\begin{table}[ht]
\caption{Default simulation parameters}
\centering
\begin{ruledtabular}
\begin{tabular}{cc}
Parameter & Value \\
\hline
$\omega_{s1}/2\pi$ & 10.75~GHz \\
$\omega_{s2}/2\pi$ & 10.60~GHz \\
$\omega_{s3}/2\pi$ & 10.48~GHz \\
$\omega_{s4}/2\pi$ & 10.35~GHz \\
$\omega_{sc}/2\pi$ & 10.54~GHz \\
$J_{12}^r/h$, $J_{23}^r/h$ & 2~MHz \\
$J_{2c}^r/h$, $J_{3c}^r/h$ & 12~MHz \\
$\gamma$ & 0.1~eV/V \\
$\alpha$ & 0.011~mV$^{-1}$\\
$A_\mu$ & 1~$\mu$eV/$\sqrt{\text{Hz}}$\\
$\sigma_{Bz}/2\pi$ & 0.05~MHz
\end{tabular}
\end{ruledtabular}
\label{tab:spin_coupler_param}
\end{table}

\section{Intra-qubit residual exchange} \label{sec:intra res exchange}

\subsection{Single-qubit gate error}

We begin by considering the effects of residual exchange within one qubit, which we refer to as intra-qubit residual exchange. To illustrate the effects of intra-qubit residual exchange, consider only the left RST qubit in Fig.~\ref{fig:direct_sq_model}(a), decoupled from the second qubit and subjected to a resonant exchange drive.
We split the RST qubit Hamiltonian into two parts: the static part $H_0$, which includes $J_{12}^r$, and the drive part $H_d(t)$, which includes the total exchange coupling $J_{12}(t)$ minus the static residual part. (Further details can be found in Appendix~\ref{apd:driving_rst}.) The total Hamiltonian is:

\begin{align}
    H(t) &= H_0 + H_d(t),\label{eq:single_q_drive}\\
    H_0 &= -\frac{1}{2}\hbar\omega_q\sigma_z,\\
    H_d(t) &= \frac{1}{2}\left[ J_{12}(t) - J_{12}^r \right] \left( \cos\Theta \sigma_x - \sin\Theta \sigma_z \right),
\end{align}
where $\omega_q = \sqrt{(\omega_{s1} - \omega_{s2})^2 + (J_{12}^r/\hbar)^2}$ is the qubit frequency, $J_{12}^r$ is the intra-qubit residual exchange coupling, and $\Theta$ is the angle between the quantization and exchange axes, i.e., $\cos\Theta = |\omega_{s1} - \omega_{s2}| / \omega_q$ and $\sin\Theta = J_{12}^r /\hbar \omega_q$.

We choose a logarithmic cosine envelope for the drive pulse~\cite{rimbach2023simple}. 
Figure~\ref{fig:direct_sq_pulse}(a) shows such a voltage pulse and the associated exchange pulse, which takes the form
\begin{equation}\label{eq:sq_exchange_drive}
    J_{12}(t) = J_{12}^r + J_d s(t) \left[ 1 + \cos\left( \omega_d t + \varphi \right) \right],
\end{equation}
where $J_d$ is the amplitude, $\omega_d$ is the carrier frequency, $\varphi$ is the carrier phase, and $s(t)$ is the cosine envelope. 
For a fixed pulse width $\tau$, we choose the envelope function as
\begin{equation} \label{eq:sq_envelope}
    s(t) = \frac{1}{2} - \frac{1}{2}\cos(\frac{2\pi t}{\tau}).
\end{equation}
While this work focuses on analytically transparent pulse shapes to elucidate the underlying mechanisms, the framework is fully compatible with quantum optimal control techniques. 

We analyze the dynamics by transforming into the rotating frame with respect to the driving frequency. When $\omega_d = \omega_q$, the effective Hamiltonian becomes
\begin{equation}\label{eq:sq_H_eff}
    H_{\text{eff}}(t) = \frac{1}{4} J_d s(t)\cos\Theta
    \left[
    \begin{array}{cc}
        0 & e^{i\varphi}  \\
        e^{-i\varphi} & 0
    \end{array}
    \right] + H_\varepsilon(t).
\end{equation}
Here $i$ is the imaginary unit, and $H_\varepsilon$ is the error matrix, formulated by
\begin{equation}\label{eq:sq_error_matrix}
    H_\varepsilon(t) = 
    \left[
    \begin{array}{cc}
        -\varepsilon_z(t) & \varepsilon_x(t) \\
        \varepsilon_x^\dagger(t) & \varepsilon_z(t)
    \end{array}
    \right],
\end{equation}
where the matrix elements $\varepsilon_x(t)$ and $\varepsilon_z(t)$ are
\begin{align}
    \varepsilon_x(t) &= \frac{1}{4} J_d s(t) \cos\Theta \left[ 2e^{-i\omega_d t} + e^{-i(2\omega_d t + \varphi)} \right],\label{eq:sq_error_matrix_ele_ex}\\
    \varepsilon_z(t) &= \frac{1}{2} J_d s(t) \sin\Theta\left[ 1 + \cos\left( \omega_d t + \varphi \right) \right].\label{eq:sq_error_matrix_ele_ez}
\end{align}

Equations~(\ref{eq:sq_H_eff})-(\ref{eq:sq_error_matrix_ele_ez}) suggest that any drive through the intra-qubit exchange leads to unwanted rotations about both the $x$ and $z$ axes, resulting pulse-dependent coherent errors. In particular, since exchange coupling is always positive, the non-zero time averaged value of the exchange drive will result in unwanted rotations around the $z$-axis, as well as a drive in the rotating frame at frequency $\omega_d$. The usual counter-rotating terms at frequency $2 \omega_d$ also appear and can create errors if the drive amplitude becomes comparable to the drive frequency (typically hundreds of MHz in RST qubits).
Similar effects have been reported in superconducting fluxonium qubits~\cite{rower2024suppressing, manucharyan2009fluxonium}, whose transition frequency between the ground and the first excited state is less than $1~\text{GHz}$. 

\subsection{Commensurate drive}
Equations~(\ref{eq:sq_error_matrix_ele_ex}) and~(\ref{eq:sq_error_matrix_ele_ez}) indicate that the error terms oscillate at integer multiples of the drive frequency. This fact suggests that the effects of the errors can be mitigated by choosing gate times at integer multiples of the driving period. 
To explore this possibility, we simulate the RST qubit formed from spins 1 and 2 undergoing a resonant drive with the parameters listed above. 
Figs.~\ref{fig:direct_sq_pulse}(b)–(d) illustrate the trajectories of the Bloch vector $\vec{r} = (r_x, r_y, r_z)$ during a $R(\pi/2, \pi/4)$ rotation from resonant drives with different gate times, where $\tau_d = 2\pi/\omega_d$ is the drive period.
Solid lines represent the actual rotations coming from the resonant single-qubit drives shown in Fig.~\ref{fig:direct_sq_pulse}(a), while dashed lines indicate the ideal trajectories. 
The fast oscillatory motion observed in all cases arises from counter-rotating effects and the presence of positive residual exchange. The $3.3\tau_d$ trajectory ends off the equator due to incomplete fast rotations.
In contrast, for pulse durations of $4\tau_d$ and $20\tau_d$, the trajectories terminate properly on the equator. This behavior at integer multiples of the drive-period occurs because the counter-rotating and positive-exchange errors, represented by $\varepsilon_x(t)$, destructively interfere with themselves at the end of each full drive-cycle. 
Unwanted phase accumulation is evident in all cases, highlighting the presence of the off-axis rotation error, described by $\varepsilon_z(t)$.

\begin{figure*}[t]
{\includegraphics[width= \textwidth]{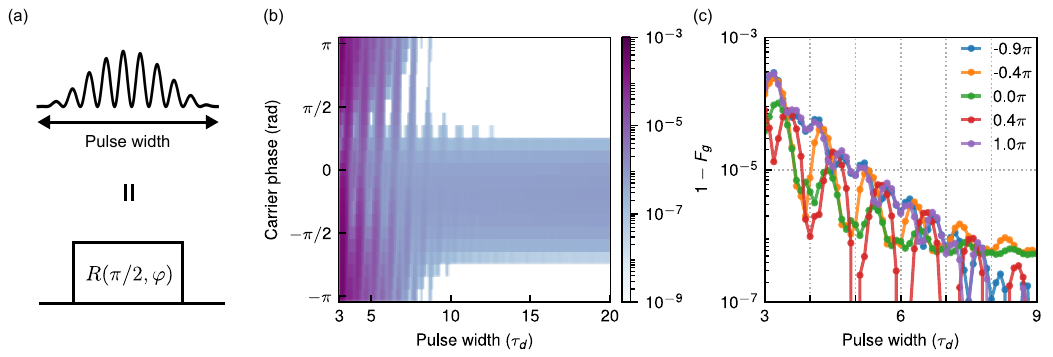}}
\caption{
Single-qubit gate for a RST qubit with intra-qubit residual exchange coupling $J_\text{intra}^r / \hbar = 2\pi \times 2~\text{MHz}$.
(a) Exchange pulse waveform with carrier phase $\varphi$ driving a qubit $\pi/2$-rotation $R(\pi/2,\varphi)$. 
(b) Coherent error of the $\pi/2$ rotation for different carrier phases and pulse widths. 
(c) Horizontal traces from panel (b) at different carrier phases. 
Local minima of the gate infidelity occur when the pulse width is an integer multiple of the driving period $\tau_d$.
}
\label{fig:direct_sq_ideal}
\end{figure*}

To explore this in more detail, we consider the unitary time evolution operator of the effective Hamiltonian in Eq.~(\ref{eq:sq_H_eff}) under an exchange pulse with pulse width $\tau$:
\begin{equation}
    U(\tau) = \mathcal{T} \exp\left[-\frac{i}{\hbar}\int_0^\tau dt~H_\text{eff}(t)\right],
\end{equation}
where $\mathcal{T}$ is the time-ordering operator. 
Expanding $U(\tau)$ using the Magnus expansion~\cite{magnus1954exponential} and retaining only the leading-order (first-order) term yields
\begin{equation}
    U^{(1)}(\tau) = \exp\left[-\frac{i}{\hbar}\int_0^\tau dt~H_\text{eff}(t)\right].
\end{equation}
We define the error integrals as
\begin{equation}
    I_{k} = \frac{1}{\hbar}\int_0^\tau dt~\varepsilon_{k}(t),
\end{equation}
where $k\in\{x,z\}$. The integral of the envelope function [Eq.~(\ref{eq:sq_envelope})] is $\int_0^\tau dt ~s(t) = \tau/2$. With these considerations, the unitary operator is then 
\begin{equation}\label{eq:drect_sq_unitary_matrix}
U^{(1)}(\tau) = \left[
    \begin{array}{cc}
        \cos\frac{\theta}{2}-i\frac{\theta_3}{\theta}\sin\frac{\theta}{2} & - \frac{\theta_2 + i\theta_1}{\theta}\sin\frac{\theta}{2} \\
        \frac{\theta_2 - i\theta_1}{\theta}\sin\frac{\theta}{2} & \cos\frac{\theta}{2}+i\frac{\theta_3}{\theta}\sin\frac{\theta}{2}
    \end{array}
    \right],
\end{equation}
where 
\begin{align}
    \theta_1 &= \frac{1}{4\hbar} J_d \tau \cos\Theta \cos\varphi + 2\Re{I_x},\\
    \theta_2 &= -\frac{1}{4\hbar} J_d \tau \cos\Theta\sin\varphi - 2\Im{I_x},\\
    \theta_3 &= -2I_z,
\end{align}
and $\theta = \sqrt{\theta_1^2 + \theta_2^2 + \theta_3^2}$. 
Notice that $U^{(1)}(\tau)$ represents an arbitrary ideal transversal rotation $R(\theta, \varphi)$ when both the error integrals approach zero. 
By doing the integrals $I_{x}$ and $I_{z}$:
\begin{equation}
\begin{aligned}
    I_x &= \frac{\pi^2 J_d\cos\Theta(1-e^{-i\omega_d \tau})}{16i\hbar\omega_d}\\
    &\times \left[ \frac{e^{-i\varphi}\left( 1 + e^{-i\omega_d\tau} \right)}{\pi^2 - \omega_d^2\tau^2} + \frac{16}{4\pi^2 - \omega_d^2 \tau^2} \right],
\end{aligned}
\end{equation}
and
\begin{equation}
\begin{aligned}
    I_z &= \frac{\pi^2 J_d \sin\Theta}{\hbar\omega_d}\\
    &\times \left[ \frac{\omega_d \tau}{4\pi^2} + \frac{\sin(\omega_d\tau + \varphi) - \sin\varphi}{4\pi^2 - \omega_d^2 \tau^2} \right],
\end{aligned}
\end{equation}
one can notice that when $\omega_d\tau = 2n\pi$ for $n\in\mathbb{Z}^+$,  $I_x = 0$ and $I_z$ become independent of $\varphi$. 
The results suggest that all coherent errors can be minimized at leading-order in the Magnus expansion by choosing the pulse duration to be an integer multiple of the driving period ($\tau_d:=2\pi/\omega_d$) $\tau = n\tau_d$, known as commensurate driving~\cite{zwanenburg2025single, rower2024suppressing}.
Notice that this technique also works when driving with a detuning, i.e., $\omega_d - \omega_q \neq 0$, where additional Z-rotations will occur and can be mitigated by modifying the virtual-Z compensations later.

While commensurate timing has been explored in other platforms, the error structure and control constraints in exchange-driven RST qubits are fundamentally different.
Previous works employ commensurate driving to suppress drive-induced non-RWA effects, such as counter-rotating contributions that vanish in the absence of driving~\cite{rower2024suppressing, zwanenburg2025single}.
In contrast, the errors considered here arise from residual always-on exchange interactions that persist even without driving.
The periodic exchange drive enables these static error terms to be averaged out over commensurate pulse durations.

At the practical level, the commensurate driving does not impose new calibration requirements beyond those already needed for high-fidelity resonant control. 
Precise knowledge of both the qubit Larmor and driving frequencies is a prerequisite for operating both Loss–DiVincenzo and singlet–triplet qubits, and slow drifts are routinely handled through standard frequency-tracking and recalibration protocols. 
The commensurate condition leverages this existing frequency knowledge through timing synchronization and does not require higher precision or real-time global clocking.

\subsection{Single-qubit gate fidelity}

To assess the reduction in single-qubit gate errors with commensurate driving, we compute an average gate fidelity based on simulations in the absence of noise (see the following section for fidelity estimates in the presence of noise and Appendix~\ref{apd:gate_fidelity} for more details). 
Although the commensurate condition is derived using a leading-order Magnus expansion, all numerical simulations are performed using the full time-dependent Hamiltonian and therefore include higher-order contributions automatically.
The $R(\pi/2,\varphi)$ gate fidelity implemented by the unitary evolution $U(\tau)$ in Eq.~(\ref{eq:drect_sq_unitary_matrix}) is
\begin{equation}
    F_g = \frac{1}{3}\left(2F_e + 1\right),
\end{equation}
where $F_e$ is the entanglement fidelity, calculated by
\begin{equation}\label{eq:SQ_RST_entang_f}
    F_e = \frac{1}{4} \sum_{\rho \in \{\rho_0\}} \text{Tr} \left[ \rho_\tau\left( \rho \right) \rho_\text{ideal}\left( \rho \right) \right].
\end{equation}
Here, $\rho_\tau(\rho) = U(\tau)\rho U^\dagger(\tau)$ and $\rho_\text{ideal}(\rho) = R(\pi/2,\varphi)\rho R(-\pi/2,\varphi)$ denote the states produced by the actual and ideal evolutions acting on the initial state $\rho$, respectively. The set of single-qubit fiducial states is  
$\{\rho_0\} = \{|0\rangle,
|1\rangle, 
\left(|0\rangle+|1\right)/\sqrt{2}, 
\left(|0\rangle+i|1\rangle\right)/\sqrt{2} 
\}$.
Figure~\ref{fig:direct_sq_ideal}(b) shows the coherent error with varying pulse width and carrier phase, where the diagrammatic pulse sequence is illustrated in Fig.~\ref{fig:direct_sq_ideal}(a), and the driving frequency is on resonance with the qubit transition frequency.
The gate error becomes lower at longer pulse widths as smaller pulse amplitudes are required. 
Figure~\ref{fig:direct_sq_ideal}(c) illustrates the gate error against pulse width with different carrier phases $\varphi$. 
The gate error reaches a local minimum around each integer multiple of the drive period $\tau_d$, consistent with our expectations for commensurate driving. 
Figure~\ref{fig:direct_sq_noise}(a) shows the coherent error against pulse width under commensurate drive with different intra-qubit residual exchange values. 
There is minimal dependence of the error on the intra-qubit residual exchange coupling, suggesting the robustness of the commensurate drive technique against residual exchange.  

\subsection{Error budget of a noisy $X_{\pi/2}$ gate}

\begin{figure}[t]
    {\includegraphics[width= 0.5\textwidth]{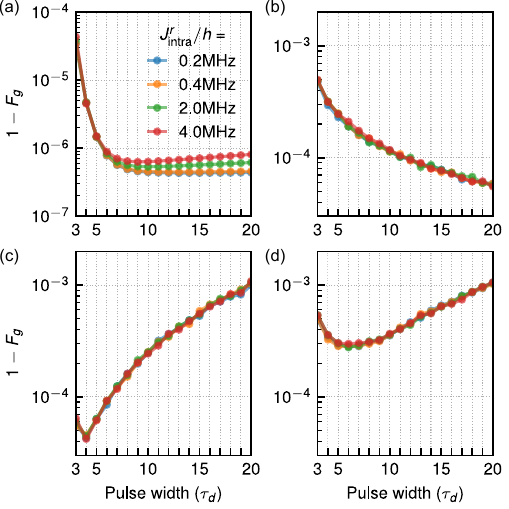}}
    \caption{
    Error budget of a commensurately-driven $\pi/2$ gate. 
    (a) Coherent error.
    (b) Charge noise error. 
    (c) Hyperfine noise error. 
    (d) All errors.
    }
    \label{fig:direct_sq_noise}
\end{figure}

To evaluate how well commensurate driving works in the presence of electrical and hyperfine fluctuations, we evaluate single-qubit $X_{\pi/2}$ gate fidelities with different noise configurations. Figs.~\ref{fig:direct_sq_noise}(b)-(d) show Monte Carlo simulations of the commensurate $X_{\pi/2}$ gate fidelity against pulse width with only charge noise, hyperfine noise, and both, respectively. 
The errors from charge noise decrease with the pulse length, because longer pulse widths require smaller exchange pulse amplitudes, reducing the sensitivity to charge noise. 
Conversely, with only hyperfine noise, the errors increase with pulse width due to increased effects of hyperfine dephasing during longer pulses.  
No significant variations in charge- or hyperfine-induced errors are observed across different values of intra-qubit residual exchange $J_{\text{intra}}^r$. 
When both charge and hyperfine noise are included, an optimum configuration around $6\tau_d$ emerges due to the competing effects of charge and hyperfine noise. 
These results suggest that, with commensurate driving, coherent errors arising from $J_{\text{intra}}^r$ can be effectively mitigated under realistic noise conditions. 

Notice that the commensurate driving protocol is compatible with other error mitigating strategies such as dynamical decoupling~\cite{bluhm2011dephasing, sun2024full} or real-time feedback~\cite{shulman2014suppressing}, which primarily mitigate low-frequency noise during idle evolution.
The commensurate suppression of all coherent errors is achieved without introducing additional control pulses or feedback loops, making the approach complementary to existing noise-mitigation strategies rather than a replacement. 

\section{Inter-qubit residual exchange} \label{sec:inter res exchange}

\subsection{Error budget of a noisy CZ gate}
We now turn to the effects of residual exchange between qubits, or inter-qubit residual exchange. As discussed above, two RST qubits interacting via inter-qubit exchange experience a ZZ-interaction, as described by Eq.~(\ref{eq:direct_zz}). This interaction naturally leads to a two-qubit CZ gate, which can be implemented by pulsing the inter-qubit exchange coupling. As discussed above, pulsing the inter-qubit exchange can lead to leakage outside the computational subspace, which can be partially mitigated by adiabatic pulse shaping~\cite{wardrop2014exchange,zhang2025universal}. 

Similar to single-qubit commensurate driving, one can also mitigate the leakage dynamics by carefully controlling the CZ gate pulse width.
The two-qubit dynamics in the presence of inter-qubit residual exchange coupling can be characterized with a time scale specific to the coupling strength to each of the leakage states. We define this timescale as
\begin{equation}\label{eq:direct_tau_cz}
\tau_{CZ} = \frac{h}{|E_{\widetilde{\uparrow\uparrow\downarrow\downarrow}} - E_{00}| + |E_{\widetilde{\downarrow\downarrow\uparrow\uparrow}} - E_{11}|}.
\end{equation}
A CZ gate implemented by a process $U_{CZ}(\tau)$ with pulse width $\tau$ is considered to be commensurately driven when the pulse width is an integer multiple of $\tau_{CZ}$.

To demonstrate this possibility in simulation, we simulate a two-qubit CZ gate and calculate the fidelity with 
\begin{equation}
F_{CZ} = \frac{1}{5}\left( 4F_{e}^{CZ}  + 1 \right),
\end{equation}
where the two-qubit entanglement fidelity is given by
\begin{equation}
F_e^{CZ} = \frac{1}{16} \sum_{\rho \in \{\rho'_0\}} \text{Tr} \left[ \rho_\tau\left( \rho \right) \rho_\text{ideal}\left( \rho \right) \right].
\end{equation}
Here $\rho_\tau(\rho) = U_{CZ}(\tau)\rho U_{CZ}^\dagger(\tau)$, $\rho_\text{ideal}(\rho) = \text{CZ}\rho \text{CZ}$ is the ideal evolution of the initial state $\rho$, and 
$\{\rho'_0\} = \{\rho_{A}\otimes \rho_{B}\}$ for 
$\rho_{A}, \rho_{B} \in \{|0\rangle,
|1\rangle, 
\left(|0\rangle+|1\rangle\right)/\sqrt{2},
\left(|0\rangle+i|1\rangle\right)/\sqrt{2} 
\}$ is a set of two-qubit fiducial states. 

Figure~\ref{fig:direct_sq_cz_noise}(a) shows the CZ gate error with charge noise versus pulse width and hyperfine noise strength. 
The non-monotonic dependence of the error on the pulse width, observed when the hyperfine noise strength exceeds $2\pi \times 0.05~\text{MHz}$, results from the dominant error mechanism shifting from leakage dynamics to hyperfine noise.
Figure~\ref{fig:direct_sq_cz_noise}(b) shows the CZ gate errors under both charge and hyperfine noise, with $\sigma_{Bz} = 2\pi \times 0.04~\text{MHz}$, plotted against pulse width and inter-qubit residual exchange coupling strength. 
No significant variation is observed across different values of inter-qubit residual exchange, suggesting the effectiveness of commensurate driving at mitigating leakage errors. 
Figure~\ref{fig:direct_sq_cz_noise}(c) shows the population in the leakage states $|\widetilde{\uparrow\uparrow\downarrow\downarrow}\rangle$, and $|\widetilde{\downarrow\downarrow\uparrow\uparrow}\rangle$, the total population outside the computational subspace, and the coherent error, all illustrated as the pulse length varies, and without noise. 
As the pulse width increases, all error sources become less significant, and gate performance eventually becomes limited primarily by hyperfine noise.
An oscillating pattern of the errors in Figure~\ref{fig:direct_sq_cz_noise}(c) between interleaved even and odd time steps appears due to residual calibration errors, which cancel out after an even number of leakage-dynamics periods.

\subsection{Single-qubit control crosstalk}

A more serious challenge associated with residual inter-qubit exchange, however, arises from the fact that the exchange coupling cannot be made zero. 
Residual inter-qubit exchange leads to a quantum crosstalk effect, which depends on the residual ZZ-interaction strength $\zeta$, and where the transition frequency of one qubit becomes dependent on the state of the other qubit. 

\begin{figure}[t]
    {\includegraphics[width= 0.5\textwidth, trim={24cm 0cm 4cm 0cm}, clip]{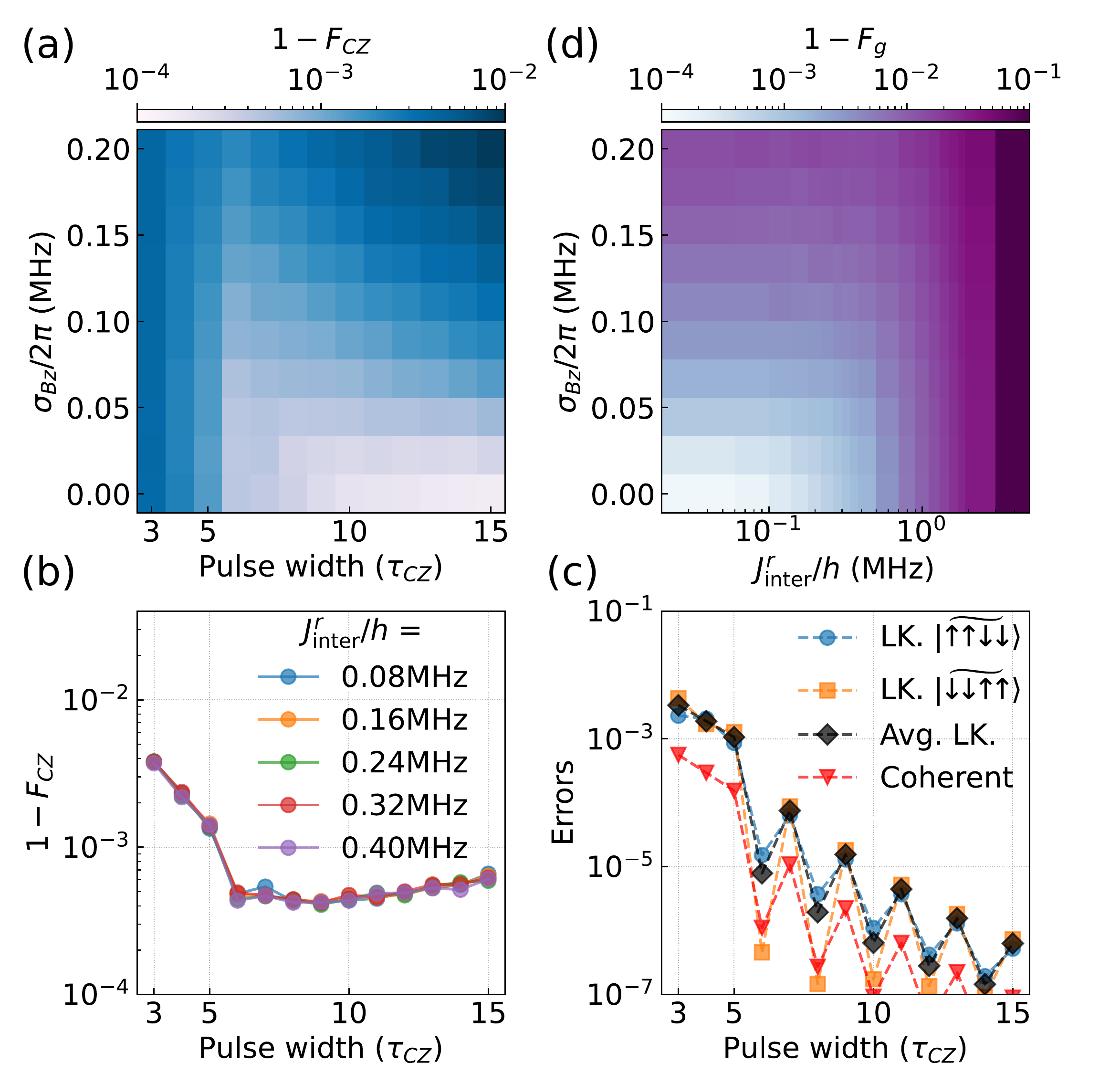}}
    \caption{
    $X_{\pi/2}$ and CZ gate error of two exchanged-coupled RST qubits.
    (a) CZ gate error with different pulse widths and hyperfine noise amplitudes with $J_{23}^r/h=0.4~\text{MHz}$.
    (b) CZ gate error with $\sigma_{Bz}=2\pi\times0.04~\text{MHz}$ and varying pulse widths and inter-qubit residual exchange values.
    (c) Leakage and coherent error budget of a CZ gate with $J_{23}^r/h=0.4~\text{MHz}$.
    (d) $X_{\pi/2}$ gate error with different inter-qubit residual exchange coupling $J_\text{inter}^r=J_{23}^r$ and hyperfine noise amplitudes. The gate time used here is $10\tau_d$.
    }
    \label{fig:direct_sq_cz_noise}
\end{figure}

To illustrate this effect, we simulate the behavior of two RST qubits as listed in Table~\ref{tab:spin_coupler_param} and assume commensurate driving. 
We focus on the single-qubit $X_{\pi/2}$ gate of the first qubit, where the system is initialized in $|\widetilde{\downarrow\uparrow\downarrow\uparrow}\rangle$. 
Figure~\ref{fig:direct_sq_cz_noise}(d) shows the single-qubit $X_{\pi/2}$ gate error on the first qubit against inter-qubit residual exchange $J_\text{inter}^r = J_{23}^r$ and hyperfine noise amplitude $\sigma_{Bz}$.
Once the residual exchange coupling exceeds the hyperfine noise amplitude, the gate error increases dramatically as a result of the crosstalk effect. 
For isotopically purified silicon, where the hyperfine noise is  $\sigma_{Bz} = 2\pi \times 0.04~\text{MHz}$~\cite{xue2022quantum}, gate errors below $1 \times 10^{-3}$, require residual exchange values less than approximately $0.12~\text{MHz}$.

\begin{figure*}[t]
{\includegraphics[width= 0.65\textwidth, trim={8cm 0.5cm 9cm 0cm}, clip]{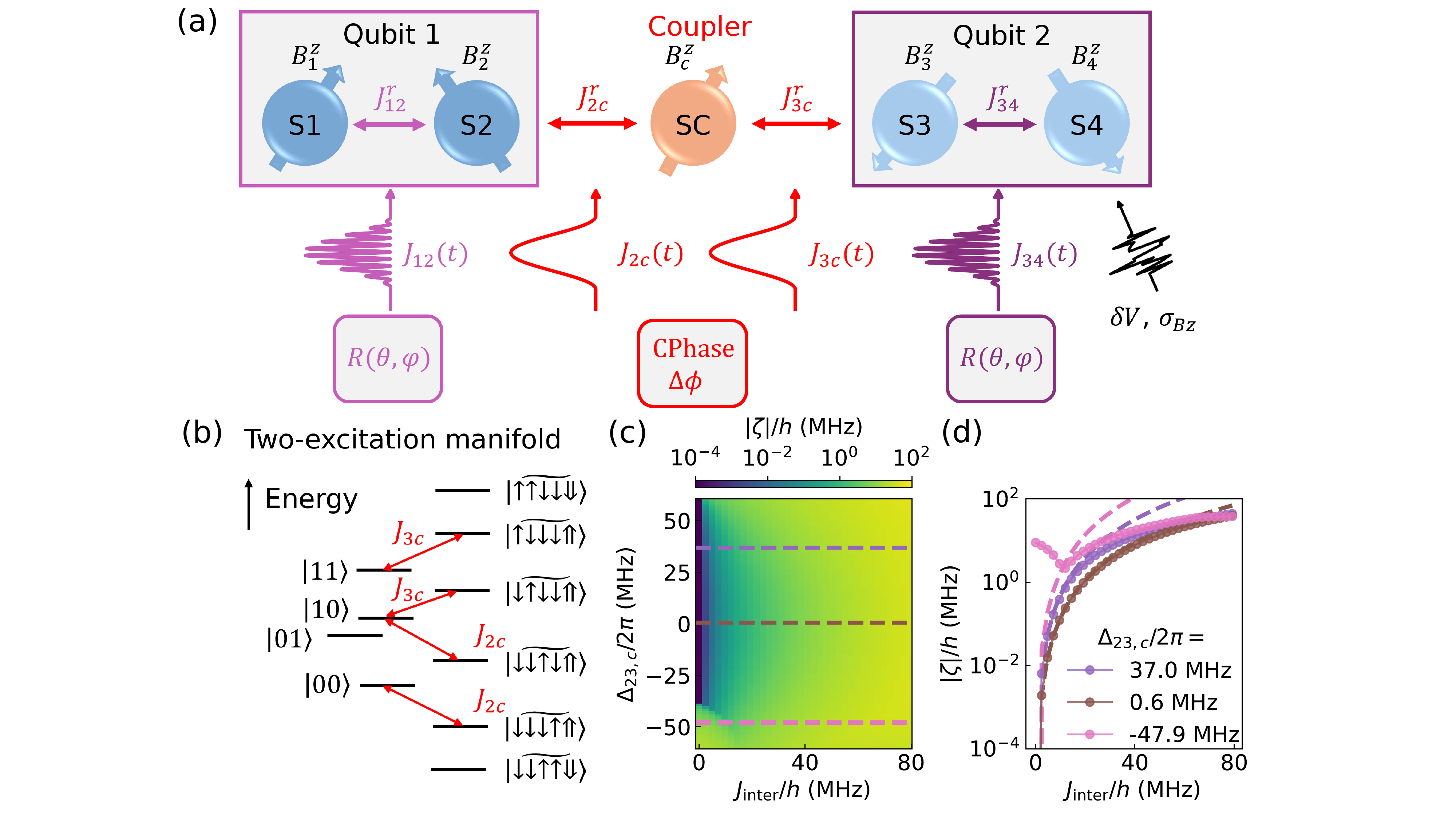}}
\caption{
(a) One-dimensional spin chain with five electrons where the middle spin serves as a coupler between two RST qubits. 
(b) Energy levels of the two-excitation manifold. The four levels on the left and the six levels on the right form the computational and leakage subspaces, respectively.
(c) ZZ-interaction $\zeta$ extracted by diagonalizing Eq.~(\ref{eq:sc_hamiltonian}) at different coupler frequencies $\omega_{sc}$ and inter-qubit residual exchange values $J_{\text{inter}}$.
The qubit-spin frequencies and intra-qubit residual exchange are as in Table~\ref{tab:spin_coupler_param}.
(d) Horizontal cuts from (c). The dashed lines are the corresponding perturbation results from Eq.~(\ref{eq:coupler_zz}). 
}
\label{fig:coupler_model}
\end{figure*}

\subsection{Single-spin coupler}

One approach to mitigating this crosstalk error is to design the device with a large dynamic range such that the exchange pulse can be sufficiently small at low gate voltages to minimize the crosstalk but simultaneously large enough at high gate voltages to implement rapid logic gates. Another approach, which we now discuss, involves introducing a single spin as a coupler between two RST qubits. 

Consider two RST qubits, each connected to a central coupler spin, as illustrated in Fig.~\ref{fig:coupler_model}(a). 
The coupler spin $S_C$ experiences a local magnetic field $B_c^z$ and is exchange-coupled to two qubit spins $S_2$ and $S_3$. 
The spin-coupler system Hamiltonian is then 
\begin{equation} \label{eq:sc_hamiltonian}
\begin{aligned}
        H_\text{sc} = &-\frac{1}{2}\sum_{k=1}^4\hbar\omega_{s,k}s_k^z + \frac{1}{4}\sum_{k=1,3}J_{k,k+1}^r \vec{s}_k\cdot\vec{s}_{k+1}\\
        &-\frac{1}{2}\hbar\omega_{s,c}s_c^z + \frac{1}{4}\sum_{k=2,3}J_{k,c}^r \vec{s}_k\cdot\vec{s}_{c},
\end{aligned}
\end{equation}
where $\hbar\omega_{s,k}=g_k\mu_BB_k^z$ is the Zeeman energy of spin $(k\in\{1,2,3,4,c\})$, $J_{12}^r$ and $J_{34}^r$ are intra-residual exchange coupling, $J_{2c}^r$ and $J_{3c}^r$ are inter-qubit residual exchange coupling.

In the remainder of the section, we consider a symmetric exchange coupling between each qubit-coupler spin pair, i.e., $J_{2c}^r=J_{3c}^r$.  
The computational subspace is defined as $\left\{ |00\rangle, |01\rangle, |10\rangle,|11\rangle \right\} = \left\{ |\widetilde{\downarrow\uparrow\downarrow\uparrow\Downarrow}\rangle, |\widetilde{\downarrow\uparrow\uparrow\downarrow\Downarrow}\rangle, |\widetilde{\uparrow\downarrow\downarrow\uparrow\Downarrow}\rangle, |\widetilde{\uparrow\downarrow\uparrow\downarrow\Downarrow}\rangle \right\}$.
The eigenstates are labeled by $|\widetilde{S_1S_2S_3S_4S_C}\rangle$, where the qubit-spin and coupler-spin are in single and double arrows, respectively.
The coupler spin remains in its ground state during normal operation.

We model our spin chain with spin frequency consecutively decreasing from $S_1$ to $S_4$, yielding an energy-level diagram inside the two-excitation manifold as shown in Fig.~\ref{fig:coupler_model}(b).
The states on the left and right sides represent the computational and leakage states, respectively.
In contrast to the scenario without a coupler, leakage errors now occur in general only when the coupler is excited. Thus, the leakage channels discussed previously no longer appear at first order because of the introduction of the coupler. 

\subsubsection{ZZ-interaction}
For large qubit-spin/coupler-spin Larmor-frequency detuning $\Delta_{kc}$, the effective ZZ-interaction between RST qubits is (see Appendix~\ref{apd:spin_coupler}):
\begin{equation}\label{eq:coupler_zz}
    \zeta \approx -\frac{J_{2c}J_{3c}}{4\hbar^2}\left( \frac{J_{2c}}{\Delta_{2c}^2} + \frac{J_{3c}}{\Delta_{3c}^2}\right),
\end{equation}
where $\Delta_{kc}=|\omega_{s,k}-\omega_{s,c}|$ for $k\in\{2,3\}$. Maximizing the on-off-ratio of the ZZ-interaction requires that $\omega_{s3}-\omega_{sc}=\omega_{sc}-\omega_{s2}$, i.e., $\Delta_{23,c}=0$, where
\begin{equation}
\Delta_{23,c} = \omega_{sc} - \frac{\omega_{s2}+\omega_{s3}}{2}.
\end{equation}
Figure~\ref{fig:coupler_model}(c) shows the ZZ-interaction for varying coupler frequency and inter-qubit exchange coupling. 
Figure~\ref{fig:coupler_model}(d) shows horizontal traces with varying coupler frequencies, where the solid and dashed lines are data extracted from (c) and the corresponding analytical solutions from Eq.~(\ref{eq:coupler_zz}), respectively. 

Since the residual ZZ-interaction arising from residual exchange coupling is quadratically suppressed by the frequency detuning between the qubit and coupler spins, a significant reduction in single-qubit gate crosstalk is expected.
Figure~\ref{fig:coupler_sq_cz_noise}(a) compares the single-qubit gate error for the direct coupling case and the midpoint-frequency coupler configuration ($\Delta_{23,c} = 0$). 
The pulse width for the single-qubit gate is set to $15\tau_d$ in this analysis.
We observe a two-order-of-magnitude reduction in coherent errors with the coupler, even at small inter-qubit residual exchange values. 
The coupler shows a tolerance of residual exchange up to $4~\text{MHz}$, where single-qubit gate errors below that point are dominated by residual coherent errors.  
Even at such high inter-qubit exchange coupling values, the ZZ-interaction reaches $2\pi \times 10~\text{MHz}$, with an on-off ratio on the order of $10^{4}$ with respect to the residual level. We expect that pulsing exchange couplings between both qubit-coupler spins to around $60~\text{MHz}$ will enable a CZ gate with a duration of tens to hundreds of nanoseconds.

\subsubsection{Error budget of noisy $X_{\pi/2}$ and CZ gates with a coupler}

\begin{figure*}[t]
    {\includegraphics[width= \textwidth, trim={0cm 0cm 0cm 0cm}, clip]{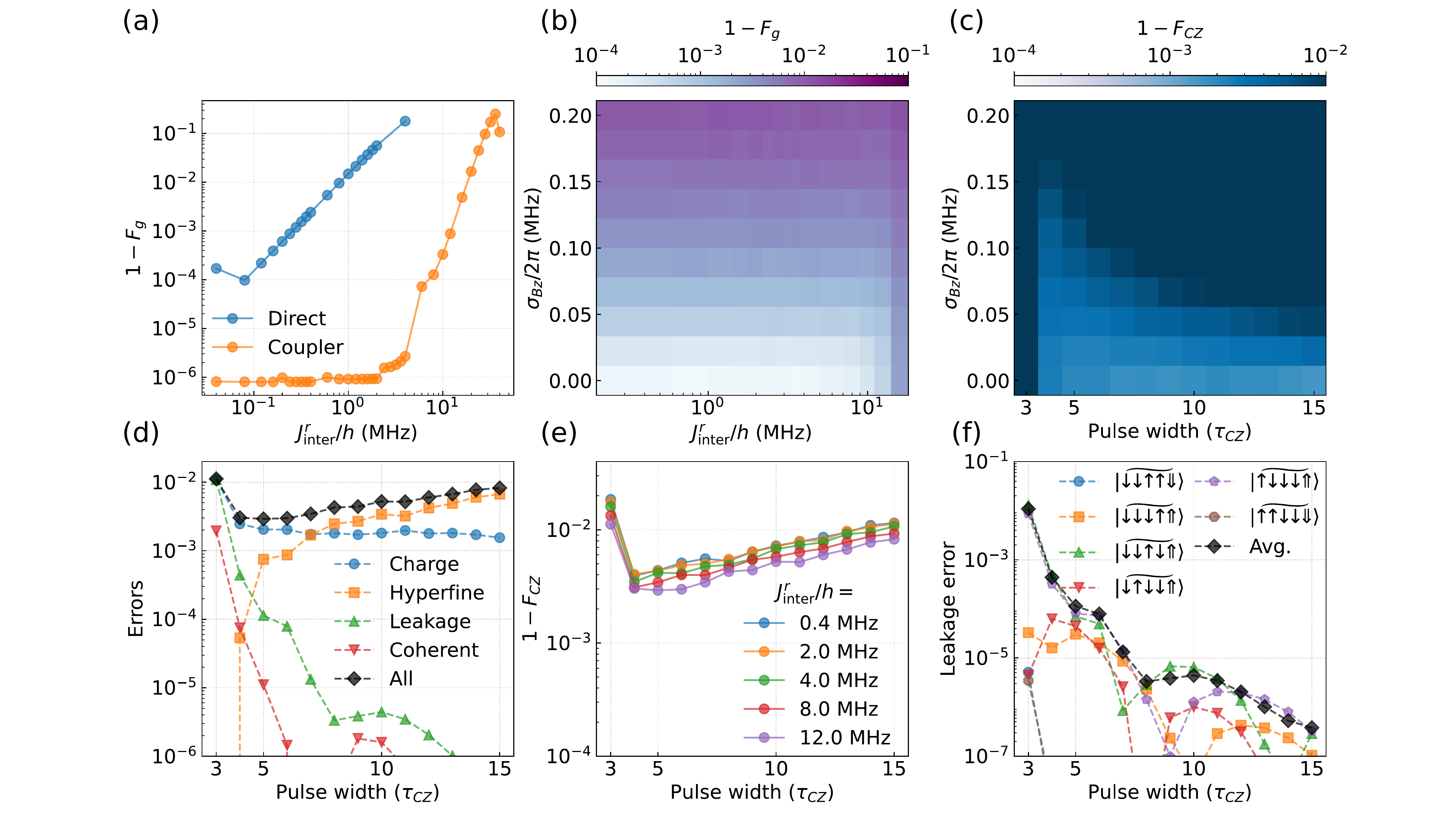}}
    \caption{
    $X_{\pi/2}$ and CZ gate error of two RST qubits coupled by a single-spin coupler.
    (a) A comparison of the coherent error of the $X_{\pi/2}$ gate against inter-qubit residual exchange between two RST qubits with direct and coupler-assisted coupling. The inter-qubit coupling for the coupler case refers to the coupler-spin couplings $J_{2c}^r=J_{3c}^r=J_\text{inter}^r$. Both cases use a gate time of $15\tau_d$, which is also used in (b). $J_\text{inter}^r/h=12~\text{MHz}$ are used in (c), (d), and (f).
    (b) $X_{\pi/2}$ gate error with varying $J_\text{inter}^r$ under different hyperfine noise strengths $\sigma_B^z$. Charge noise contributions are included.
    (c) CZ gate error with varying CZ gate pulse width and hyperfine noise strength. Charge noise contributions are included.
    (d) Error budget of the CZ gate with hyperfine noise strength $\sigma_{Bz}=2\pi\times0.04~\text{MHz}$.
    Here, the optimum near $4\tau_{CZ}$ pulse width reflects the trade-off between leakage, charge noise, and hyperfine errors for the parameter of Tab.~\ref{tab:spin_coupler_param}.
    (e) CZ gate error with varying CZ gate pulse width and inter-qubit residual exchange coupling. Both noise types are applied at the same level as (d). 
    (f) Leakage error of a coupler-assisted CZ gate with varying CZ gate pulse width. 
    }
    \label{fig:coupler_sq_cz_noise}
\end{figure*}

In the following discussion we assume the coupler can be perfectly initialized on its ground state. The single-qubit $X_{\pi/2}$ gate error with varying inter-qubit residual exchange $J_\text{inter}^r$ and hyperfine noise values $\sigma_{Bz}$ is illustrated in Fig.~\ref{fig:coupler_sq_cz_noise}(b).
In contrast to the results shown in Figure~\ref{fig:direct_sq_cz_noise}, the coherent error when $\sigma_{Bz}=2\pi\times 0.04~\text{MHz}$ remains small up to residual exchange values up to around $10~\text{MHz}$. This robustness to residual exchange illustrates the potential of the spin coupler. 

Turning to the CZ gate error, we set a large inter-qubit residual exchange value such that $J_\text{inter}^r/h=12~\text{MHz}$ for estimating the gate performance under the worst-case scenario.
In this regime, which corresponds to the largest $J_\text{inter}^r/h$ such that the single-qubit gate error is no longer dominated by inter-qubit crosstalk, the exchange-to-charge noise sensitivity remains at a high value, which contributes to a large charge noise error background. 
Figure~\ref{fig:coupler_sq_cz_noise}(c) shows the coupler-assisted CZ gate error plotted against the CZ gate pulse width and hyperfine noise level. 
Similar to the directly-coupled case in Eq.~(\ref{eq:direct_tau_cz}), the unit cycle time $\tau_{CZ}$ for commensurate driving is based on the inverse of the average energy spacing between each of the coupled computational-leakage state pairs: 
\begin{equation}
\begin{aligned}
    \tau_{CZ}^{-1} = \frac{1}{4 h}\left( |\Delta E_{\widetilde{\uparrow\downarrow\downarrow\downarrow\Uparrow},11}| + |\Delta E_{\widetilde{\downarrow\downarrow\downarrow\uparrow\Uparrow},00}| \right. \\
    \left. + |\Delta E_{\widetilde{\downarrow\uparrow\downarrow\downarrow\Uparrow},10}|+ |\Delta E_{\widetilde{\downarrow\downarrow\uparrow\downarrow\Uparrow},10}| \right),
\end{aligned}
\end{equation}
where $\Delta E_{kl}=E_k - E_l$ is the energy difference between states $|k\rangle$ and $|l\rangle$.
As the coupler frequency is fixed at $\Delta_{23,c}=0$ and the system is configured with the parameters listed in Table~\ref{tab:spin_coupler_param}, we calculate $\tau_{CZ}\approx16.50~\text{ns}$. 

For hyperfine noise levels lower than $2\pi\times 0.50~\text{MHz}$, charge noise becomes dominant error source and contributes an error background of about $2\times10^{-3}$. 
As above, the non-monotonic dependence of the CZ gate error on the pulse width occurs due to the tradeoff between leakage and hyperfine errors.  
The error budget with a fixed hyperfine noise amplitude $\sigma_{Bz} = 2\pi \times 0.04~\text{MHz}$ is shown in Fig.~\ref{fig:coupler_sq_cz_noise}(d), where the optimum configuration is at about $4\tau_{CZ}\approx 66~\text{ns}$, corresponding to a CZ gate error of $3 \times 10^{-3}$.
At short pulse widths, both coherent leakage into the coupler-excited states~[Fig.~\ref{fig:coupler_sq_cz_noise}(f)] and charge noise errors, both of which are about an order of magnitude larger than the residual coherent error, dominate the gate error~[Fig.~\ref{fig:coupler_sq_cz_noise}(d)]. 
As the pulse width increases, the charge noise contribution decreases as smaller exchange amplitudes are required, reducing the charge-noise sensitivity.
Since the coupler-assisted CZ gate requires a higher-amplitude inter-qubit exchange pulse compared to the directly-coupled case, an overall increase in the charge noise contribution is observed compared to the directly-coupled case.
The leakage error also decreases with longer pulse widths due increased envelope adiabaticity, and the gate error is limited primarily by hyperfine dephasing. 
The error budget shown here is obtained by isolating individual error mechanisms and should be interpreted as a diagnostic comparison rather than a strictly additive decomposition of the full quantum error channels. 
Figure~\ref{fig:coupler_sq_cz_noise}(e) shows the CZ gate error with different pulse widths and inter-qubit residual exchange values in the presence of both hyperfine and charge noise at the same level of Fig.~\ref{fig:coupler_sq_cz_noise}(d). 
We observe a small dependence of the error on $J_\text{inter}^r$ due to the small variations of charge noise sensitivities. 

The primary source of leakage arises from coupler excitations due to the low adiabaticity of the exchange pulse envelope. 
By shaping the envelope as a cosine function without a flat plateau, the total leakage error can be suppressed below $10^{-4}$. 
Figure~\ref{fig:coupler_sq_cz_noise}(f) shows the leakage error during a CZ gate. 
At the trade-off point of $4\tau_{CZ}$, the leakage error is approximately $4 \times 10^{-4}$—significantly lower than the contributions from both charge and hyperfine noise.
The observed state dependence of the leakage arises from asymmetric coupling of different computational basis states to non-computational levels under the exchange pulse, but remains subdominant across all regimes considered.
Incoherent leakage arising from relaxation processes corresponds to erasure errors and is expected to be negligible on the gate timescales considered in our simulations~\cite{borjans2019single}.

As mentioned earlier, this analysis assumes perfect initialization of the coupler, for which an optimal error rate of $2.6 \times 10^{-3}$ has been realistically demonstrated for single-spin initialization~\cite{mills2022high}.
Any residual coupler excitation will result in a leakage (erasure-type) error, which is a challenge also encountered in superconducting qubits with tunable couplers~\cite{sung2021realization, yang2024coupler}. 
These erasure errors do not coherently propagate within the computational subspace and therefore contribute additively to the overall gate error, which is expected to be small compared to the dominant dephasing and slow exchange-noise-induced errors.

Fast and high-fidelity coupler initialization is therefore essential for scalable operation of spin-coupler architectures. 
In addition to high initialization fidelity, active-reset can also reduce state preparation errors. 
By entangling the coupler spin with an auxiliary spin in a neighboring dot and reading out the auxiliary spin via Pauli-spin-blockade-based parity measurement, one could apply conditional single-spin control pulses to deterministically drive the coupler into its ground state.
This protocol would run on the same timescale as, and concurrently with, qubit initialization. 
While the error rate above reflects single-spin initialization benchmarks, this specific auxiliary-spin active-reset protocol can achieve a comparable initialization error. 
Measurement durations as short as $2.4~\mu$s with initialization errors of $2\times10^{-3}$ have been demonstrated in silicon double quantum dots~\cite{takeda2024rapid}.
Because coupler reset would occur once per shot prior to gate operation, it would add to the existing per-shot initialization overhead without extending the sub-100 ns gate time itself.

A complementary approach for checking the coupler spin mid-circuit, without collapsing an ongoing computation, is quantum non-demolition (QND) measurement of the coupler spin~\cite{xue2020repetitive}. We leave the integration of QND-based mid-circuit coupler monitoring, and the associated overhead, as a direction for future work.

\section{Outlook and conclusion}

The RST–coupler–RST architecture offers a pathway to constructing crosstalk-free unit cells for large-scale semiconductor quantum dot arrays.
It preserves compatibility with large-scale quantum error correction and shuttling-based approaches. 
The coupler mediates interactions only during active gate operations and remains effectively decoupled during idle periods, ensuring that repeated QEC cycles and transport operations are not disrupted. 
Because the commensurate timing constraint is local to individual gates, it does not impose global synchronization across the processor and can be naturally integrated into surface-code-like architectures.

In conclusion, we analyzed single- and two-qubit control errors resulting from both inter- and intra-qubit residual exchange coupling in systems of exchange-coupled RST qubits. 
We showed that commensurate driving can mitigate coherent and leakage errors in both single- and two-qubit gates.
For intra-qubit residual exchange values below a few hundred kHz, gate fidelities are limited by the interplay of hyperfine and charge noise. For residual exchange values above this level, single-qubit gates suffer from a quantum crosstalk error. We showed that a single-spin coupler between RST qubits can suppress this error by two orders of magnitude. The large on-off ratio that can be achieved with the coupler suggests that a fast CPhase gate with minimal coherent and leakage error can be achieved.

\section{Data Availability}
The data and simulation scripts are available at https://doi.org/10.5281/zenodo.17794872~\cite{duan_2025_17794872}.

\section{Acknowledgments}
J.D. acknowledges discussions with Pan Shi, Meng Wang, and Feiyang Ye. 
This work was sponsored by the Army Research Office through Grant No. W911NF-23-1-0115 and the Air Force Office of Scientific Research through Grant No. FA9550-23-1-0710. The views and conclusions contained in this document are those of the authors and should not be interpreted as representing the official policies, either expressed or implied, of the Army Research Office or the U.S. Government. The U.S. Government is authorized to reproduce and distribute reprints for Government purposes notwithstanding any copyright notation herein.

\appendix
\section{Driving a RST qubit}\label{apd:driving_rst}

In this section we will derive the single-RST-qubit Hamiltonian under a resonant intra-qubit exchange drive with the rotating wave approximation.
Considering a Heisenberg model for two spins with corresponding Zeeman terms, the total Hamiltonian is given by
\begin{equation}
    H_\text{s} = -\frac{1}{2}\hbar\omega_{s1}s_1^z - \frac{1}{2}\hbar\omega_{s2}s_2^z + \frac{1}{4} J^r_{12} \vec{s}_1 \cdot \vec{s}_2,
\end{equation}
where $\hbar$ is the reduced Planck constant, $J_{12}^r$ is the residual exchange between spins $S_1$ and $S_2$,  $s^\mu_k$ with $\mu\in\{x,y,z\}$ and $k=1,2$ are the Pauli operators for spin $k$. The Pauli-vector $\vec{s}_k = (s_k^x, s_k^y, s_k^z)$, and the Larmor frequency for spin $k$ is defined as $\hbar\omega_{s,k} = g_k \mu_B B_k^z$.  

We construct the encoded Pauli operators in the two-spin decoherence-free subspace ($\text{DFS}_2$) as
\begin{align}
    \bar{\sigma}_x &= \frac{1}{2}\left( s_1^x s_2^x + s_1^y s_2^y \right),\\
    \bar{\sigma}_y &= \frac{1}{2}\left( s_1^y s_2^x - s_1^x s_2^y \right),\\
    \bar{\sigma}_z &= -\frac{i}{2}\left[ \bar{\sigma}_x , \bar{\sigma}_y \right].
\end{align}
By ignoring a global phase depending on $J^r_{12}$, the Hamiltonian within the $\text{DFS}_2$ is given by
\begin{equation}\label{eq:sq_static_H}
    \bar{H}_0 = -\frac{1}{2}\hbar\bar{\omega}\bar{\sigma}_z + \frac{1}{2} J_{12}^r \bar{\sigma}_x,
\end{equation}
where $\bar{\omega} = |\omega_{s1} - \omega_{s2}|$.
After diagonalization, the RST qubit static Hamiltonian becomes
\begin{equation}
    H_0 = -\frac{1}{2} \hbar\omega_q \sigma_z,
\end{equation}
where $\omega_q = \sqrt{\bar{\omega}^2 + (J_{12}^r/\hbar)^2}$, and $\sigma_\mu$, $\mu\in\{x,y,z\}$, are the Pauli operators of the RST qubits. 
The unitary transformation between basis $\sigma_k$ and $\bar{\sigma}_k$ is a rotation operator along the $y$-axis
\begin{equation}\label{eq:SQ_unitary_trans_oper}
    \bar{R}_y\left(\Theta \right) = \exp\left( -i\frac{\Theta}{2} \bar{\sigma}_y \right),
\end{equation}
where $\cos\Theta = \bar{\omega}/\omega_q$, and $\sin\Theta = J_{12}^r/(\hbar\omega_q)$.
When an intra-qubit exchange drive $J_{12}(t)$ is applied, the time-dependent Hamiltonian can be formulated into static [Eq.~(\ref{eq:sq_static_H})] and driven parts, given by
\begin{equation}
\begin{aligned}
    \bar{H}(t) &= -\frac{1}{2}\hbar\bar{\omega}\bar{\sigma}_z + J_{12}(t)\bar{\sigma}_x \\
    &=\bar{H}_0 + \bar{H}_d(t). \\
\end{aligned}
\end{equation}
with 
\begin{align}
    \bar{H}_d(t) &= \frac{1}{2} \left[ J_{12}(t) - J_{12}^r \right]\bar{\sigma}_x.
\end{align}
Note that while $J_{12}(t)$ contains $J_{12}^r$, the driven part contains only the time-dependent exchange coupling. 

Moving into the qubit eigenbasis by applying the unitary transformation in Eq.~(\ref{eq:SQ_unitary_trans_oper}), the Hamiltonian $H(t)=\bar{R}_y\left(\Theta \right)\bar{H}(t)\bar{R}^\dagger_y\left(\Theta \right)$ becomes
\begin{align}
    H(t) &= H_0 + H_d(t), \\
    H_d(t) & = \frac{1}{2} \left[ J_{12}(t) - J_{12}^r \right] \left( \cos\Theta\sigma_x - \sin\Theta \sigma_z \right).
\end{align}
The exchange driven single-qubit rotations have both $x$ and $z$ components, due to the non-zero intra-qubit residual exchange coupling. 
By taking the drive as in Eq.~(\ref{eq:sq_exchange_drive}), the effective Hamiltonian in the rotating frame with respect to the driving frequency becomes
\begin{equation}
    \begin{aligned}
        H_{\text{eff}}(t) &= i\hbar\dot{U}_{\text{rt}}U_{\text{rt}}^\dagger + U_{\text{rt}}HU_{\text{rt}}^\dagger \\
        &= \tilde{H}_0 + \tilde{H}_d(t),
    \end{aligned}
\end{equation}
where $U_\text{rt} = \exp\left( -i\omega_d t \sigma_z/2 \right)$. 
The static and driving terms in the rotating frame are given by
\begin{align}
    \tilde{H}_0 &= -\frac{1}{2}\hbar\delta_q\sigma_z, \\
    \tilde{H}_d(t) &= \frac{1}{2} \left[ J_{12}(t) - J_{12}^r \right] \notag \\
        &\quad \times\left( \cos\Theta\, U_\text{rt}\sigma_xU_\text{rt}^\dagger - \sin\Theta\, U_\text{rt}\sigma_zU_\text{rt}^\dagger \right),
\end{align}
where $\delta_q = |\omega_q - \omega_d|$.  The transformed Pauli operators are given by
\begin{align}
    U_\text{rt}\sigma_xU_\text{rt}^\dagger &= \left[
    \begin{array}{cc}
         0 & e^{-i\omega_d t} \\
         e^{i\omega_d t} & 0 
    \end{array}
    \right],\\
    U_\text{rt}\sigma_z U_\text{rt}^\dagger &= \sigma_z.
\end{align}
When the drive frequency is resonant with the qubit frequency, i.e., $\delta_q=0$, the effective Hamiltonian in the rotating frame reduces to Eq.~(\ref{eq:sq_H_eff}).

\section{Direct coupling}\label{apd:direct_coupling}

In this section, we will calculate the ZZ-interaction from residual inter-qubit exchange coupling. 
Consider a one-dimensional spin chain with spins at each site labeled as $S_k$, where $k \in {1, 2, 3, 4}$.
The Hamiltonian is 
\begin{equation}
    H_\text{s} = -\frac{1}{2}\sum_{k=1}^4\hbar\omega_{s,k}s_k^z + \frac{1}{4}\sum_{k=1}^3J_{k,k+1}^r \vec{s}_k\cdot\vec{s}_{k+1},
\end{equation}
where $\hbar\omega_{s,k}$ is the Zeeman energy of spin $k$, and $J_{k,k+1}^r$ is the residual exchange coupling between spin $k$ and $k+1$.
For small residual exchange coupling ($J_{k,k+1}\ll \hbar|\omega_{s,k}-\omega_{s,k+1}|$), the Heisenberg part can be treated as a perturbation.
The perturbative expansion of the eigenenergies up to second-order is given by
\begin{equation}
    \begin{aligned}
        E_n &= E_n^{(0)} + \sum_{k}\frac{J_{k,k+1}^r}{4}E_{n,k}^{(1)}+\sum_{k,l}\frac{J_{k,k+1}^rJ_{l,l+1}^r}{16}E_{n,kl}^{(2)},
    \end{aligned}
\end{equation}
where the correction at each order is calculated as
\begin{align}
    E_{n,k}^{(1)} &= \langle n^{(0)}|\vec{s}_k\cdot\vec{s}_{k+1}|n^{(0)}\rangle,\\
    E_{n,kl}^{(2)} &= \sum_{m\neq n} \frac{\langle n^{(0)}|\vec{s}_k\cdot\vec{s}_{k+1}|m^{(0)}\rangle\langle m^{(0)}|\vec{s}_l\cdot\vec{s}_{l+1}|n^{(0)}\rangle}{E^{(0)}_n-E^{(0)}_m}.
\end{align}
Here $|n^{(0)}\rangle\in \mathcal{C}$ is the unperturbed eigenstates inside the computational subspace $\mathcal{C}$, $|m^{(0)}\rangle \in \mathcal{M}$ is unperturbed eigenstates inside the two-excitation manifold $\mathcal{M}$, and $k,l\in\{1,2,3\}$.
The set of unperturbed eigenstates with two excitations is $\mathcal{M} = \mathcal{C} \cup \mathcal{L}$, where the basis states for the computational and leakage subspaces are $\mathcal{C} = \left\{ |\downarrow\uparrow\downarrow\uparrow\rangle, |\downarrow\uparrow\uparrow\downarrow\rangle, |\uparrow\downarrow\downarrow\uparrow\rangle, |\uparrow\downarrow\uparrow\downarrow\rangle \right\}$, and $\mathcal{L} = \left\{  |\uparrow\uparrow\downarrow\downarrow\rangle, |\downarrow\downarrow\uparrow\uparrow\rangle \right\}$, respectively.
The static ZZ-interaction is defined as
\begin{equation}
    \zeta = E_{\uparrow\downarrow\uparrow\downarrow} + E_{\downarrow\uparrow\downarrow\uparrow} - E_{\uparrow\downarrow\downarrow\uparrow} - E_{\downarrow\uparrow\uparrow\downarrow},
\end{equation}
which is identical to Eq.~(\ref{eq:direct_zz}).
The only non-zero contribution from the perturbative expansion is the first-order correction
\begin{equation}
\begin{aligned}
        \zeta^{(1)} &= \frac{1}{4}\sum_{k=1}^3 J_{k,k+1}^r\left(E_{\uparrow\downarrow\uparrow\downarrow,k}^{(1)} + E_{\downarrow\uparrow\downarrow\uparrow,k}^{(1)} \right.\\
        &\left.- E_{\uparrow\downarrow\downarrow\uparrow,k}^{(1)} - E_{\downarrow\uparrow\uparrow\downarrow,k}^{(1)}\right)\\
        & = -J_{23}^r.
\end{aligned}
\end{equation}
Therefore, the residual ZZ-interaction is given by the first-order contribution $\zeta^{(1)}$
\begin{equation}
    \zeta = -J_{23}^r.
\end{equation}

\section{Spin coupler} \label{apd:spin_coupler}
In this section we calculate the ZZ-interaction mediated by a single-spin coupler between two RST qubits.
Consider a one-dimensional Heisenberg model for five spins with Hamiltonian
\begin{equation}
\begin{aligned}
        H_\text{sc} = &-\frac{1}{2}\sum_{k=1}^4\hbar\omega_{s,k}s_k^z + \frac{1}{4}\sum_{k=1,3}J_{k,k+1}^r \vec{s}_k\cdot\vec{s}_{k+1}\\
        &-\frac{1}{2}\hbar\omega_{sc}s_c^z + \frac{1}{4}\sum_{k=2,3}J_{k,c}^r \vec{s}_k\cdot\vec{s}_{c},
\end{aligned}
\end{equation}
where $\hbar\omega_{s,k}=g_k\mu_BB_k^z$ is the Zeeman energy of spin $k\in\{1,2,3,4,c\}$, $J_{12}^r$ and $J_{34}^r$ intra-qubit residual exchange couplings, $J_{2c}^r$ and $J_{3c}^r$ are inter-qubit residual exchange couplings.
Treating all diagonal terms ($s_k^z$ and $s_k^z s_l^z$) as static $H_0$ and all $x$ and $y$ terms as perturbations, we define
\begin{equation}
        V = \frac{1}{2}\sum_{k=1,3}J_{k,k+1}^r V_{k,k+1}+\frac{1}{2}\sum_{l=2,3}J_{l,c}^r V_{l,c},
\end{equation}
where $V_{k,k+1} = s^+_k s^-_{k+1} + s^-_k s^+_{k+1} $, and $V_{k,c} = s^+_k s^-_{c} + s^-_k s^+_{c}$ for $k\in\left\{1,3\right\}$ and $l\in\left\{2,3\right\}$, respectively.

Using labels $|S_1S_2S_3S_4S_c\rangle$ to denote states, the basis states for the unperturbed computational and leakage states are $\mathcal{C}=\left\{ |\downarrow\uparrow\downarrow\uparrow\Downarrow\rangle, |\downarrow\uparrow\uparrow\downarrow\Downarrow\rangle, |\uparrow\downarrow\downarrow\uparrow\Downarrow\rangle, |\uparrow\downarrow\uparrow\downarrow\Downarrow\rangle \right\}$, and $\mathcal{L}=\left\{ |\uparrow\downarrow\downarrow\downarrow\Uparrow, \downarrow\uparrow\downarrow\downarrow\Uparrow\rangle, |\downarrow\downarrow\uparrow\downarrow\Uparrow\rangle, |\downarrow\downarrow\downarrow\uparrow\Uparrow\rangle, |\downarrow\downarrow\uparrow\uparrow\Downarrow\rangle, |\uparrow\uparrow\downarrow\downarrow\Downarrow\rangle \right\}$. For clarity, we use the double arrow to represent the coupler spin. 

The ZZ-interaction is given by
\begin{equation}
    \zeta = E_{\uparrow\downarrow\uparrow\downarrow\Downarrow}+E_{\downarrow\uparrow\downarrow\uparrow\Downarrow} - E_{\uparrow\downarrow\downarrow\uparrow\Downarrow}-E_{\downarrow\uparrow\uparrow\downarrow\Downarrow}.
\end{equation}
The terms giving non-zero contribution to this ZZ-interaction appear at second-order
\begin{equation}
    \zeta^{(2)} = E_{\uparrow\downarrow\uparrow\downarrow\Downarrow}^{(2)}+E_{\downarrow\uparrow\downarrow\uparrow\Downarrow}^{(2)} - E_{\uparrow\downarrow\downarrow\uparrow\Downarrow}^{(2)}-E_{\downarrow\uparrow\uparrow\downarrow\Downarrow}^{(2)},
\end{equation}
where the second-order energy correction is calculated by
\begin{equation}
    E_n^{(2)} = \sum_{m\neq n}\frac{\langle n^{(0)}|V|m^{(0)}\rangle\langle m^{(0)}|V|n^{(0)}\rangle}{E_n^{(0)} - E_m^{(0)}}.
\end{equation}
Here $|n^{(0)}\rangle\in\mathcal{C}$ and $|m^{(0)}\rangle\in\mathcal{C}\cup\mathcal{L}$ are the unperturbed eigenstates, and $E_n^{(0)}$ is the unperturbed eigenenergy.
The final result shows
\begin{equation}
\begin{aligned}
    \zeta^{(2)} = &\frac{(J_{2c}^r)^2}{2} \left( \frac{1}{2\hbar\Delta_{2c} +J_{3c}^r - J_{12}^r } \right.\\
    &\left.- \frac{1}{2\hbar\Delta_{2c} - J_{3c}^r - J_{12}^r } \right)\\
    &+\frac{(J_{3c}^r)^2}{2} \left( \frac{1}{2\hbar\Delta_{3c} +J_{2c}^r - J_{34}^r } \right.\\
    &\left.- \frac{1}{2\hbar\Delta_{3c} - J_{2c}^r - J_{34}^r } \right),
\end{aligned}
\end{equation}
where $\Delta_{kl} = \omega_{s,k}-\omega_{s,l}$ for $k,l\in\{1,2,3,4,c\}$. 
When the nearest-neighbor detuning is large, i.e. when $2\hbar\Delta_{2c} \gg J_{3c}^r\pm J_{12}^r$ and $2\hbar\Delta_{3c}\gg J_{3c}^r\pm J_{12}^r$, the ZZ-interaction can be expanded to
\begin{equation}
    \zeta \approx -\frac{J_{2c}^rJ_{3c}^r}{4\hbar^2}\left( \frac{J_{2c}^r}{\Delta_{2c}^2} + \frac{J_{3c}^r}{\Delta_{3c}^2} \right).
\end{equation}

\section{Maximum gate voltage} \label{apd:max_voltage}
\begin{figure*}[t]
    {\includegraphics[width=0.72\textwidth]{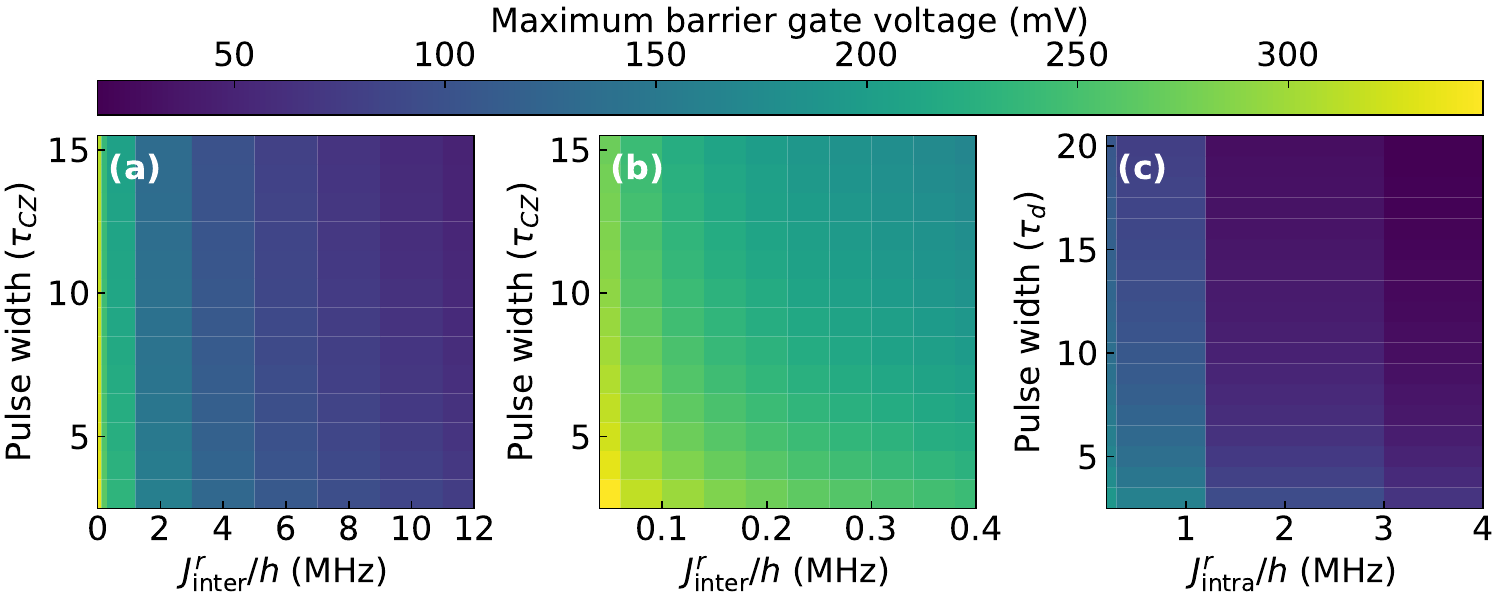}}
    \caption{
    Maximum barrier gate voltage required to implement calibrated exchange pulses for the parameters listed in Table~I. 
    (a) Coupler-assisted CZ gate as a function of pulse width and inter-qubit residual exchange $J_{\mathrm{inter}}^r$. 
    (b) Directly coupled CZ gate under the same conditions. 
    (c) Single-qubit $X_{\pi/2}$ gate of Q1 as a function of pulse width and intra-qubit residual exchange $J_{\mathrm{intra}}^r$. 
    The color scale indicates the maximum barrier gate voltage reached during the exchange pulse. 
    For small residual exchange, both CZ implementations require relatively large barrier voltages, while in the large residual-exchange regime the coupler-assisted architecture achieves the required interaction with substantially reduced gate voltage compared to the directly coupled case. 
    Single-qubit gates generally require significantly smaller voltages across all regimes considered.
    }
    \label{fig:gate_voltage}
\end{figure*}

In experimental implementations of exchange-based spin qubits, the maximum barrier gate voltage is a critical constraint, as over-biasing can lead to permanent tuning shifts or damage to the control electrodes. 
It is therefore important to quantify the peak gate voltages required by the exchange pulses used in our simulations.
Figure~\ref{fig:gate_voltage} shows the maximum barrier gate voltage reached during calibrated exchange pulses for (a) the coupler-assisted CZ gate, (b) the directly coupled CZ gate, and (c) the single-qubit $X_{\pi/2}$ gate, under the parameters listed in Table~\ref{tab:spin_coupler_param}. 
For small inter-qubit residual exchange, both CZ implementations require relatively large barrier voltages to achieve the desired interaction strength. 
In contrast, in the regime of large residual exchange, the coupler-assisted architecture significantly reduces the required barrier gate voltage compared to the directly coupled case, consistent with the suppression of static ZZ-interaction. 
Single-qubit gates generally require substantially smaller gate voltages across all parameter regimes considered. 
These results indicate that, in addition to mitigating crosstalk, the coupler-assisted architecture relaxes voltage constraints for two-qubit gates in experimentally relevant regimes.

\section{Charge noise} \label{apd:charge_noise}
We simulate the effect of charge and hyperfine noise using a Monte Carlo approach, in which each result is averaged over 512 realizations.
To extract information about the individual RST qubits, we compute the partial traces of the overall density matrix. 
We generate a simulated charge noise time series $\delta V[n]$ using the method mentioned in Refs.~\cite{yang2019achieving,kawakami2016gate}. 
For a simulated evolution from time $0$ to $T$ taking $N$ points, a zero mean and unit variance Gaussian white noise time series $u[n]\sim \mathcal{N}(0,1)$ with length $N$ is generated.
The Fourier transform of $u[n]$ is computed as $\tilde{u}[n]=\mathcal{F}\{u\}[n]$.
For a noise time series $\delta V[n]$ with $1/f^\beta$ power spectrum density (PSD)
\begin{equation}
S_{V_B}(f) = \frac{A_\mu^2/\gamma^2}{f^\beta},
\end{equation}
its Fourier component $\delta \tilde{V}[n]$ can be computed as
\begin{equation}
    \delta\tilde{V}[n] = \frac{A_\mu}{\gamma}\sqrt{\frac{N}{T}\left(\frac{n - 1}{T}\right)^{-\beta}} \tilde{u}[n],
\end{equation}
where $A_\mu$ is the amplitude of the power spectral density at one Hertz, and $\gamma=0.1~\text{eV}/\text{V}$ is the lever arm.
The time series of the noise is obtained by the inverse Fourier transform of the Fourier coefficients:
\begin{equation}
    \delta V[n] = \mathcal{F}^{-1}\{\delta \tilde{V} \}[n].
\end{equation}

In the $n$-th point during a single-run of the simulation, the applied barrier voltage at this point is given by $V_B[n] + \delta V[n]$.
Here, $V_B[n]$ is the barrier gate voltage resulting a desired exchange coupling $J[n]$ at point $n$ of during exchange drive $J(t)$.
Using the empirical relation between exchange coupling and barrier voltage in Eq.~(\ref{eq:voltage_2_exchange}), the noisy exchange coupling series can be calculated through
\begin{equation}
    J[n] = J^r e^{2\alpha \left( V_B[n] + \delta V[n] \right)},
\end{equation}
where $J^r$ is the residual exchange coupling, and $\alpha$ is the barrier-exchange lever arm. 
During each run of the simulation, the charge noise series is regenerated based on this method.

\section{Hyperfine noise}\label{apd:hyperfine_noise}
\begin{figure}[t]
    {\includegraphics[width= 0.5\textwidth, trim={14cm 0cm 13cm 8cm}, clip]{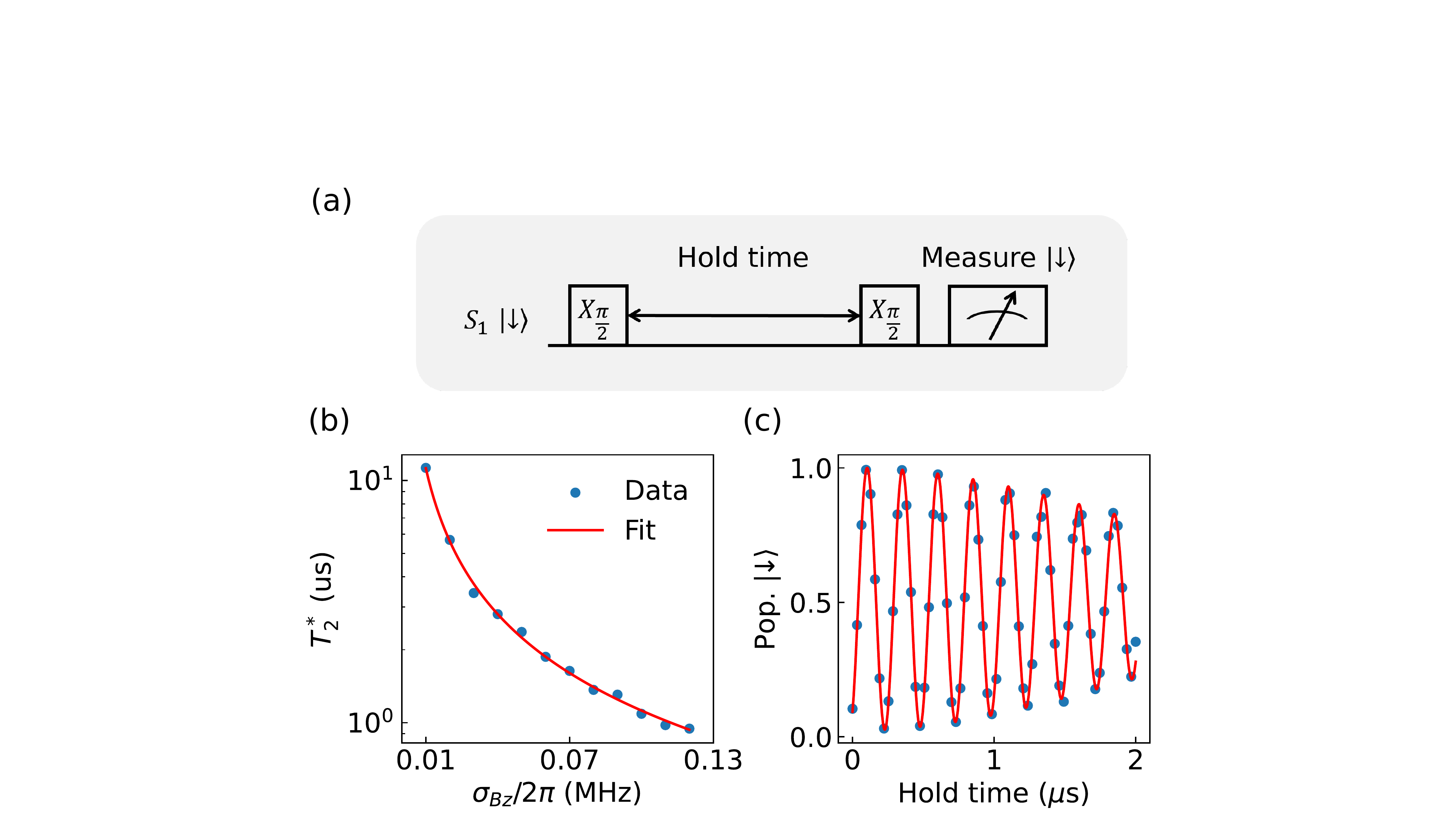}}
    \caption{
    Simulating a single-spin (Loss-Divincenzo) qubit coherence time under varying hyperfine noise levels.
    (a) Ramsey sequence for measuring $T_2^*$.
    (b) Extracted $T_2^*$ fitted with the quasistatic model.
    (c) Ramsey data with hyperfine noise level $\sigma_{Bz}=2\pi\times0.04$MHz with fitted $T_2^* = 2.81\mu$s.
    }
    \label{fig:hyperfine_coherence}
\end{figure}

A quasistatic white hyperfine noise is applied on individual spins. 
For a single spin in isotopically purified $^{28}$Si, the coherence time limited by hyperfine noise is above $5~\mu$s \cite{neyens2024probing, steinacker2025industry, xue2022quantum}.
The coherence time limited by the hyperfine noise amplitude in the simulation is calibrated by a simulated Ramsey sequence as shown in Fig.~\ref{fig:hyperfine_coherence}(a).
The simulated $T_2^*$ at different values $\sigma_{Bz}$ is fitted by a quasistatic noise model:
\begin{equation}
    T_2^*(\sigma_{Bz}) = \frac{\Omega}{\sigma_{Bz}},
\end{equation}
where the fitted result is shown Fig.~\ref{fig:hyperfine_coherence}(b) with $\Omega = 0.112$.
We extract the $T_2^*$ by fitting the simulated data (spin-down population $P_{\downarrow}$ vs hold time $t$) with a Gaussian decay envelope
\begin{equation}
    P_{\downarrow}(t) = A e^{-\left(t/T_2^*\right)^2}\cos\left( 2\pi f t + \phi \right) + B.
\end{equation}
Figure~\ref{fig:hyperfine_coherence}(c) show a simulated Ramsey experiment with $\sigma_{Bz} = 2\pi\times0.04$~MHz, where the fitted $T_2^* = 2.81~\mu$s. 

\section{Correlated noise} \label{apd:correlated_noise}

In this section, we evaluate the robustness of commensurately driven gates against correlated charge noise in the single-spin coupler regime. 
As discussed in the main text, practical implementations of coupler-based RST qubits typically require an enhanced spin–orbit coupling, which is commonly achieved by using on-chip micromagnets. 
While such strong spin-orbit coupling enables electrical control of individual spin rotations, it can also convert charge noise into effective magnetic-field fluctuations that contribute to spin decoherence. 
If multiple electron spins couple to the same charge-noise source, correlated fluctuations can arise simultaneously in both exchange interactions and Zeeman energies.

\begin{figure}[t]
    {\includegraphics[width=0.3\textwidth, trim={0cm 0cm 0cm 0cm}, clip]{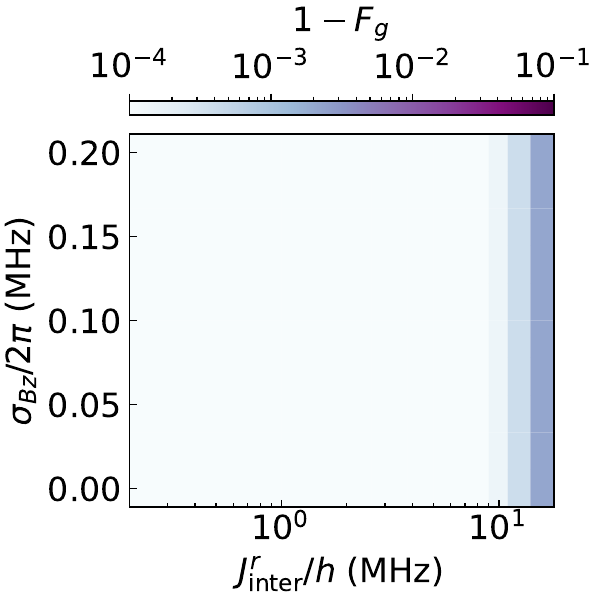}}
    \caption{
    Simulated commensurate single-qubit $X_{\pi/2}$ gate infidelity in the single-spin coupler regime under correlated exchange and magnetic-field noise. 
    Here $\sigma_{Bz}$ denotes the strength of the correlated Zeeman energy noise.
    }
    \label{fig:correlated_sq_gate_error}
\end{figure}

Using the same system configuration as in Figure~\ref{fig:coupler_sq_cz_noise}, we consider an extreme scenario in which all exchange and magnetic-field fluctuations acting on the spins originate from a common noise source. 
To model this situation, we replace the uncorrelated quasi-static hyperfine noise used in the main text with quasi-static correlated Zeeman energy noise that shares the same random seed as the exchange noise affecting all spins. 
Under this correlated noise model, we evaluate the commensurate single-qubit $X_{\pi/2}$ gate error as a function of the correlated Zeeman energy noise strength $\sigma_{Bz}$ and the inter-qubit residual exchange coupling $J_\text{inter}^r$ [Fig.~\ref{fig:correlated_sq_gate_error}].

The results show an extremely weak dependence of the gate error on the correlated Zeeman energy noise affecting all spins. 
This behavior arises because the RST qubit frequency depends primarily on the \emph{difference} between Zeeman energies of the two spins forming the qubit, which is largely insensitive to common-mode fluctuations. 
Since uncorrelated quasi-static hyperfine noise is not included in this simulation, the overall gate error is lower than the results shown in Fig.~\ref{fig:coupler_sq_cz_noise}(b). 
Instead, the dominant error sources in this regime arise from inter-qubit crosstalk and charge-noise-induced fluctuations in the exchange interaction. 
Therefore, the contribution of correlated charge noise to the commensurate single-qubit gate error is expected to be a minor effect.

From a practical perspective, the introduction of a spin coupler increases the spatial separation between data qubits by a distance exceeding the size of a single quantum dot. 
This increased spacing may further suppress the influence of correlated charge noise, whose correlation length is estimated to be on the order of $100~\mathrm{nm}$~\cite{rojas2026scaling}. 
Consequently, the coupler serves not only as a tunable interaction element but also as an architectural tool that reshapes both stochastic-noise-induced and residual-exchange-induced error channels.

\section{Gate fidelity} \label{apd:gate_fidelity}

The averaged gate fidelity $F_g$ \cite{poyatos1997complete, nielsen2002simple, pedersen2007fidelity} is used to evaluate both single- and two-qubit gate fidelities in our system. 
A straightforward approach to calculate the average gate fidelity of a noisy process $U(\tau)$ from time $0$ to $\tau$ compared to a noiseless ideal process $U_\text{ideal}$ is from its entanglement fidelity, given by
\begin{equation}
    F_e = \frac{1}{4^d} \sum_{\rho \in \{\rho_0\}} \text{Tr} \left[ \rho_\tau\left( \rho \right) \rho_\text{ideal}\left( \rho \right) \right].
\end{equation}
Here, $\{ \rho_0\}$ is a complete set of fiducial states in a $n-$qubit Hilbert space, $\rho_\tau(\rho) = U(\tau)\rho U^\dagger(\tau)$, and $\rho_\text{ideal}(\rho) = U_\text{ideal}\rho U_\text{ideal}^\dagger$.
The relation between the average gate fidelity $F_g$ and entanglement fidelity $F_e$ under this $n-$qubit Hilbert space is given by \cite{harvey2018coupling, kang2025remote}
\begin{equation}
    F_g = \frac{2^n F_e + 1}{2^n+1}.
\end{equation}

For single-qubit $n=1$ case, the complete set of fiducial states is given by
\begin{equation}\label{eq:single_qubit_entan_fidelity}
    \left\{\rho_0\right\} = \left\{|0\rangle, |1\rangle, \frac{1}{\sqrt{2}}\left(|0\rangle + |1\rangle\right), \frac{1}{\sqrt{2}}\left(|0\rangle + i|1\rangle\right)\right\}.
\end{equation}
Inside the single-RST-qubit computational subspace $\left\{|0\rangle,|1\rangle\right\}=\left\{|\widetilde{\downarrow\uparrow}\rangle, |\widetilde{\uparrow\downarrow}\rangle \right\}$, Eq.(\ref{eq:single_qubit_entan_fidelity}) becomes exactly the set of initial states in Eq.(\ref{eq:SQ_RST_entang_f}).
The single-qubit entanglement $F_e$ and averaged gate fidelity $F_g$ are given by
\begin{align}
    F_e &= \frac{1}{4} \sum_{\rho \in \{\rho_0\}} \text{Tr} \left[ \rho_\tau\left( \rho \right) \rho_\text{ideal}\left( \rho \right) \right],\\
    F_g &= \frac{1}{3} \left( 2F_e + 1 \right).
\end{align}

For the two qubit $n=2$ case, the complete set of fiducial states is given by
\begin{equation}\label{eq:two_qubit_entan_fidelity}
    \left\{\rho'_0\right\} = \left\{ \rho_A \otimes \rho_B| \rho_A, \rho_B \in \{\rho_0\} \right\}.
\end{equation}
For the two-qubit CZ gate, its entanglement $F_{CZ}^e$ and averaged gate fidelity $F_{CZ}$ are given by
\begin{align}
    F_{CZ}^e &= \frac{1}{16} \sum_{\rho \in \{\rho'_0\}} \text{Tr} \left[ \rho_\tau\left( \rho \right) \rho_\text{ideal}\left( \rho \right) \right],\\
    F_{CZ} &= \frac{1}{5} \left( 4F_{CZ}^e + 1 \right),
\end{align}
where $U_\text{ideal} = \text{diag}(1,1,1,-1)$.

\end{document}